\documentclass[useAMS,usenatbib]{mn2e}

\title[Cooling in spiral shocks]
{The ISM in spiral galaxies: can cooling in spiral shocks produce molecular clouds?}
\author[C. L. Dobbs, S. C. O Glover, P. C. Clark, R. S. Klessen]
{C. L. Dobbs,$^1$\thanks{E-mail:
dobbs@astro.ex.ac.uk} S. C. O. Glover,$^2$ P. C. Clark$^3$ \& R. S. Klessen$^3$ \\
$^1$ School of Physics, University of Exeter, 
Stocker Road, Exeter, EX4 4QL \\
$^2$ Astrophysikalisches Institut Potsdam, An der Sternwarte 16, D-14482, Potsdam, Germany \\
$^3$ Institut f\"ur Theoretische Astrophysik, Universit\"at Heidelberg, Albert-Ueberle-Str. 2, Heidelberg, Germany} 
\usepackage{amssymb}
\usepackage{amsmath}
\usepackage{graphicx}
\usepackage{epsfig}

\def\msp{{M$_{\odot}$} pc$^{-2}$ }
\def\mspnospace{{M$_{\odot}$} pc$^{-2}$}

\begin{document}
\date{\today}

\pagerange{\pageref{firstpage}--\pageref{lastpage}} \pubyear{0000}

\maketitle

\label{firstpage}

\begin{abstract} We investigate the thermodynamics of the interstellar medium (ISM) and the formation of
molecular hydrogen through numerical simulations of spiral galaxies. The model follows the chemical,
thermal and dynamical response of the disc to an external spiral potential. Self-gravity and magnetic
fields are not included. The calculations demonstrate that gas can cool rapidly when subject to a spiral
shock. Molecular clouds in the spiral arms arise through a combination of compression of the ISM by the
spiral shock and orbit crowding. These results highlight that local self-gravity is not required to form
molecular clouds. Self-shielding provides a sharp transition density, below which gas is essentially
atomic, and above which the molecular gas fraction is $>0.001$. The timescale for gas to move between
these regimes is very rapid ($\leq$1 Myr). From this stage, the majority of gas generally takes between
10 to 20 Myr to obtain high H$_{2}$ fractions ($>50$ percent). These are however strict upper limits to
the H$_{2}$ formation timescale, since our calculations are unable to resolve turbulent motions on
scales smaller than the spiral arm, and do not include self-gravity. True cloud formation timescales are
therefore expected to be even shorter. 

The mass budget of the disc is dominated by cold gas residing in the spiral arms. Between 50 and 75
percent of this gas is in the atomic phase. When this gas leaves the spiral arm and drops below the
self-shielding limit it is heated by the galactic radiation field. Consequently, most of the volume in
the interarm regions is filled with warm atomic gas. However, some cold spurs and clumps can survive in
interarm regions for periods comparable to the interarm passage timescale. Altogether between 7 and 40
percent of the gas in our disc is molecular, depending on the surface density of the calculation, 
with approximately 20\% molecular for a surface density comparable to the solar neighbourhood.
\end{abstract}

\begin{keywords}
galaxies: spiral -- hydrodynamics -- ISM: clouds -- ISM: molecules 
-- stars: formation -- galaxies:structure
\end{keywords}

\section{Introduction}
Most stars form in giant molecular clouds (GMCs), and so understanding how
GMCs form and disperse is a major goal of  
star formation research 
in a Galactic context. GMC formation is likely to be strongly dependent on the nature of the interstellar medium (ISM). Two prominent theories that have previously been advanced are that GMCs form by large scale magnetic and/or gravitational instabilities in thermally or turbulently supported gas \citep{Elmegreen1979,Balbus1985,Elmegreen1994,Chou2000,Kim2006}, or  
by the collisional build up of smaller clouds or clumps (e.g. \citealt{FieldS1965,Levinson1981,Tomisaka1984,Roberts1987,Kwan1987}). The difference between these two scenarios is that in the first the ISM behaves essentially as a fluid, whilst in the latter the behaviour is  more ballistic. The first case resembles the warm, low density component of the ISM, whilst the second 
appears more consistent 
with the cold HI or even molecular component which is predominantly situated in clumps. In the majority of galaxies, the ISM may be expected to consist of both of these components, with both instabilities and collisions contributing to GMC formation \citep{Zhang1999,Dobbs2008}. 

A major limitation of many galactic simulations is that the ISM is assumed to be isothermal, either warm \citep{Kim2002,Slyz2003,Chak2003,Li2005b,Li2006,Shetty2006}
cold HI \citep{Dobbs2006}, or a two phase medium \citep{Dobbs2007}. In this paper we include a comprehensive treatment of the thermodynamics of the ISM in global simulations, to investigate whether cooling in spiral shocks produces a clumpy, molecular phase even in the absence of self gravity. The only other calculations of heating and cooling in grand design galaxies are presented in the recent paper by \citet{Wada2008}, who uses an energy equation. Here it is possible to see how the nature of the shock changes with time as the gas changes from a warm smooth distribution to cold clumps. A similar approach has been adopted for simulations without a spiral potential, which show complex thermal and density distributions \citep{Wada2001,Wada2002,Tasker2006}. 

Another key issue related to GMC formation is the determination of the timescale over which these clouds are formed. Although there is no generally accepted definition of the formation timescale, one plausible measure is the time
required for the gas to become fully molecular. To address this, simulations ideally need to include the chemistry of H$_2$. The formation of H$_2$ has recently been modelled in simulations of dwarf galaxies \citep{Pelupessy2006}, but only a very simplistic treatment has been applied to spiral galaxies \citep{Wada2000,DBP2006}.
  
Simulations which include a treatment of the thermodynamics of the ISM typically focus on much smaller length scales than those we study here (see, for instance, the recent review by \citealt{Henne2007}).  Simulations of shock compressed layers show the formation of structure in the cold neutral medium (CNM), and the generation of low levels of turbulence \citep{Koyama2002,Heitsch2005,Vaz2006,HAudit2007}. Even without self gravity, densities in excess of 1000~cm$^{-3}$ are produced \citep{Audit2005}. The inclusion of self gravity leads to star formation in the density enhancements from the initial turbulence \citep{Vaz2007}. 
In addition to the thermodynamics, \citet{Glover2007b} also follow the evolution of molecular hydrogen, They find that in a turbulent environment, H$_2$ formation takes place over timescales of a few Myr, suggesting that molecular
clouds may form very quickly.

In previous work \citep{DBP2006}, we showed that molecular clouds can form from cold gas by agglomeration in a spiral shock \citep{Dobbs2008}. However these isothermal calculations assumed that either cold gas is prevalent in the ISM, or that warm gas cools very quickly in spiral shocks. Here we investigate whether those calculations are plausible, by including cooling and heating of the ISM. We include the thermodynamics described in \citet{Glover2007a}, which takes into account the main processes responsible for cooling and heating of the ISM, rather than assuming a prescribed relationship between density and temperature.
There are essentially two possible outcomes. 
If cooling is very rapid in the spiral shock, cold clumps are likely to form and the agglomeration of these clumps into GMCs is then a possibility. In this scenario, we are also interested to see whether the gas remains cold in the interarm regions, and whether spurs can be produced. Alternatively, cooling may not be particularly efficient, and little structure is produced in the spiral arms. In this case, another mechanism would be required to form GMCs. 

\section{Calculations}
We perform these calculations using smoothed particle hydrodynamics (SPH), a Lagrangian fluids code. The code is based on an original version by Benz \citep{Benz1990}, but has since been subject to significant modifications. The most substantial changes include the addition of sink particles \citep{Bate1995} and magnetic fields \citep{Price2005,Price2007}, although they are not used in the current paper. The code also includes individual timesteps and variable smoothing lengths. The density and smoothing lengths are solved iteratively according to 
\begin{equation}
h=\nu \bigg(\frac{m}{\rho}\bigg)^{1/3}
\end{equation}
\citep{PM2007}, where $h$ is the smoothing length, $\rho$ the density, $m$ the mass of the 
particle, and $\nu$ is a dimensionless parameter set to 1.2 in order that each particle has  $\sim$60 neighbours. Artificial viscosity is included to treat shocks, using the standard parameters, $\alpha=1$ and $\beta=2$ \citep{Monaghan1985}.

\subsection{Galactic potential and initial distribution}
We model a gaseous disc, and represent the stellar component of the galaxy by an external potential. The external potential includes a stellar disc, halo and incorporates a 4 armed spiral component.
The potential is described in full in \citet{Dobbs2006}.  The spiral component is from \citet{Cox2002} and has a pattern speed of
$2 \times 10^{-8}$~rad~yr$^{-1}$, and a pitch angle of $15^o$. The amplitude of the stellar spiral perturbation is $1.1 \times 10^{12}$ cm$^2$ s$^{-2}$. 

Particles are initially distributed randomly in a torus with radii 5 kpc $<r<$ 10 kpc. The gas particles are assigned circular velocities according to the disc potential, and in addition, a Gaussian velocity dispersion of 6 km s$^{-1}$ is imposed (which constitutes the $z$ component of the velocities). The spiral perturbation emerges with time as the simulation progresses. Initially all the gas has a temperature of 7000 K and a scale height of 400 pc. 
We perform 3 simulations, with surface densities of 4, 10 and 20 \mspnospace. Although most galaxies do not have average surface densities as low as 4~\mspnospace, the lowest density calculation would correspond to the outer regions of galaxies. The total surface density in our calculations also includes helium, which has an abundance (by number) one-tenth that of hydrogen, i.e. $n({\rm He}) = 0.1 n$ where $n({\rm He})$ is the number density of helium atoms and $n$ is the number density of hydrogen nuclei (in all forms).

We neglect both self gravity and magnetic fields in these calculations. These are
discussed in previous calculations using an isothermal equation of state \citep{Kim2002,Shetty2006,DP2008,Dobbs2008}. Rather, we are interested in whether spiral shocks alone are sufficient to cool gas to temperatures of $\leq 100$ K and produce significant amounts of molecular gas. In the next sections we describe the H$_2$ chemistry in these simulations, and outline the main thermodynamic processes included.

\subsection{Formation of H$_2$}
\label{h2_form}
The abundance of molecular gas is computed according to the expression given in \citet{Bergin2004}, 
\begin{equation}
\frac{dn({\rm H}_2)}{dt}=R_{gr}(T) n\thinspace
n({\rm H})-[\zeta_{\rm cr}+\zeta_{\rm diss}(N({\rm H_2}),A_V)]n({\rm H_2}),
\end{equation} 
where $n({\rm H})$ and $n({\rm H_{2}})$ 
denote the number densities of H and H$_2$, respectively, $n=n({\rm H})+2n(\rm H_{2})$ is the total number density of hydrogen nuclei,
$N({\rm H_{2}})$ is the column density of molecular hydrogen, $A_V$ is the visual extinction 
and $T$ is the temperature. The formation rate of H$_2$ on grains is $R_{gr}(T) n\thinspace n({\rm H})$,  where $R_{gr}=2.2 \times 10^{-18} S T^{0.5}$~cm$^{3}$~s$^{-1}$. The formation rate assumes that the number density of dust grains is proportional to the total density \citep{Holl1971}, with the ratio of the
dust grain number density to the hydrogen nuclei number density $n$ being incorporated in $R_{gr}$. The parameter $S$ is the efficiency of H$_2$ formation on grains and is assumed to be a constant value of 0.3. In practice, $S$ depends on both the gas temperature and the grain temperature \citep{Holl1989,Cazaux2004}. At high temperatures, this is unimportant, as the density of the warm gas in our simulations is in any case too low to prevent immediate dissociation of almost all of the H$_2$ formed on the grains \citep{DBP2006}. In cold gas, on the other hand, we may underestimate the H$_2$ formation rate somewhat.

The H$_2$ is primarily destroyed by photodissociation. The photodissociation rate in these calculations is given by 
\begin{equation}
\zeta_{\rm diss}(N({\rm H}_2), A_V) = f_{\rm shield}(N(H_2)) f_{\rm dust} \zeta_{\rm diss}(0)
\end{equation}
\citep{Draine1996},  where $\zeta_{\rm diss}(0)$ is the photodissociation rate for unshielded H$_2$, $f_{\rm shield}$ is a factor 
accounting for the effects of H$_2$ self-shielding, and $f_{\rm dust}$ is a factor accounting for the effects
of dust absorption. For the unshielded photodissociation rate, we take $\zeta_{diss}(0)=4.17\times10^{-11}$~s$^{-1}$,
as appropriate for an ultraviolet radiation field equivalent to the galactic average from B0 stars. For 
$f_{\rm shield}$, we use the following fitting function from \citet{Draine1996}
\begin{equation}
\begin{split}
f_{\rm shield}(N({\rm H}_2))&=\frac{0.965}{(1+x/3)^2}+\frac{0.035}{(1+x)^{0.5}} \\
&\times \exp[-8.5\times10^{-4}(1+x)^{0.5}], \\
\end{split}
\end{equation}
where $x = N({\rm H_{2}}) / 5 \times 10^{14} \: {\rm cm^{-2}}$, and where we have assumed
a typical Doppler broadening parameter $b = 3 \: {\rm km} \: {\rm s^{-1}}$. For $f_{\rm dust}$, we again follow
\citet{Draine1996} and assume that
\begin{equation}
f_{\rm dust} = e^{-\tau_{d,1000}},
\end{equation}
where $\tau_{d,1000}$ is the optical depth due to dust at a wavelength $\lambda = 1000$~\AA, given by
$\tau_{d,1000} = 2 \times 10^{-21} (N_{\rm H} + 2 N_{\rm H_{2}})$. For gas with $R_V \equiv A_V / E(B - V)
= 3.1$, as is typical of the diffuse ISM,  $\tau_{d,1000}$ and the visual extinction $A_V$ are related by
$\tau_{d,1000} = 3.74 A_V$.

In addition to photodissociation, H$_2$ is subject to destruction by cosmic rays. We assume a constant cosmic ray 
ionisation rate of $\zeta_{cr}=6 \times 10^{-18}$ s$^{-1}$, although this term is minimal compared to the 
photodissocation rate, unless $N({\rm H}_2)$ or $A_V$ are very large.

During the simulation, the fraction of molecular gas is always non-zero; the H$_2$ is never completely photodissociated. 
When the H$_2$ fraction is very small, the photodissociation rate is large, and the requirement that $n({\rm H}_2)+\Delta t_{\rm chem} \times dn({\rm H}_2)/dt>0$
forces us to use a chemical timestep $\Delta t_{\rm chem}$ that is small compared to the other timescales in the problem. To  
accommodate this without compromising the overall efficiency of the code, we subcycle the evolution of the H$_2$ abundance, splitting
each hydrodynamical timestep into as many chemical substeps as are required (see also \citealt{Glover2007a}). The abundance of H$_2$ is used to determine the cooling rate, and so is consistently coupled with the evolution of the energy during the simulation.

\subsubsection{Calculating the H and H$_2$ column densities}
The H$_2$ photodissociation rate is strongly coupled to both the H$_2$ column density (through $f_{\rm shield}$) and
the total column density of hydrogen nuclei, $N_{\rm tot} = N({\rm H}) + 2 N({\rm H_{2}})$ (through $f_{\rm dust}$).
To calculate these column densities, we use the simple estimate that the column density (of H$_2$ or hydrogen
nuclei, as appropriate) is just the local density times the typical distance to a B0 star, a constant length scale, 
$l_{\rm ph}$. The average distance to a B star is likely to be between 20 and 50 pc depending on which spectral types are included \citep{Mihalas1981,Reed2000,Maiz2001}. We therefore test 3 different length scales in these simulations. Our fiducial value is $l_{\rm ph}$ = 35 pc, but we perform calculations with 15 and 100 pc, to estimate the minimum and maximum amounts of 
H$_2$ that may be produced. In previous calculations \citep{DBP2006,Dobbs2008} we took $l_{\rm ph}$ = 100 pc.  


\subsection{Thermodynamics}
To follow the thermal evolution of the gas in our simulations, we use a model for the heating and cooling of atomic
and molecular gas taken from \citet{Glover2007a}. This model incorporates cooling from the fine structure emission lines 
of ${\rm C^{+}}$, ${\rm O}$, and ${\rm Si^{+}}$, ro-vibrational emission from ${\rm H_{2}}$, gas-grain energy transfer,
and recombination on grain surfaces. In hot gas, additional cooling comes from the collisional dissociation of 
${\rm H_{2}}$, the collisional ionisation of atomic hydrogen, and emission from atomic resonance lines 
(e.g.\ Lyman-$\alpha$) and bremsstrahlung. Heating is provided primarily by photoelectric emission from dust grains,
with additional contributions coming from cosmic ray ionisation, ${\rm H_{2}}$ photodissociation, and the pumping
of excited vibrational states of ${\rm H_{2}}$ by UV excitation or during the formation process of the molecules.
Full details of the rates adopted for these various processes can be found in \citet{Glover2007a}.

In dense gas, dust extinction significantly reduces the photoelectric heating rate. We follow \citet{Bergin2004}
and assume that the heating rate scales with the visual extinction $A_V$ as 
\begin{equation}
\Gamma_{\rm pe}(A_V) = \Gamma_{\rm pe}(0) e^{-2.5 A_{V}},
\end{equation}
where $\Gamma_{\rm pe}(0)$ is the photoelectric heating rate in optically thin gas, and where $A_V$ is
calculated as in section~\ref{h2_form} above.

Aside from our simplified treatment of the dust extinction, the main approximation that we have made is 
our neglect of the effects of emission from atomic carbon and silicon (which are not tracked in our 
chemical model) and from CO. This simplification means that the temperature of the dense, highly molecular 
gas in our simulations will not be computed entirely correctly. However, at the densities probed by our 
current simulations, the errors involved should be small, since ${\rm C^{+}}$ cooling is comparable in 
effectiveness to CO cooling, as the significant quantities of cold gas produced in our simulations 
readily demonstrate (see e.g.\ section~\ref{thermal_distrib}; see also \citealt{GloverJ2007}). Moreover, we would expect the dynamics 
of this cold, dense component to be dominated by turbulent motions, rendering its precise temperature
of limited importance.

\section{Results}
\begin{table}
\centering
\begin{tabular}{c|c|c|c|c}
 \hline
Model & $\Sigma$ & $l_{\rm ph}$ & No. particles & M$_{\rm part}$  \\
 & (\msp) & (pc) & & (M$_{\odot}$) \\
\hline
1 & 4 & 35 & 8 $\times 10^6$ & 125 \\
2 & 10 & 35 & 8 $\times 10^6$ & 312 \\
3 & 20 & 35 & 8 $\times 10^6$ & 625 \\
4 & 10 & 15 & 8 $\times 10^6$ & 312 \\
5 & 10 & 100 & 8 $\times 10^6$ & 312 \\
6 & 10 & 35 & 1 $\times 10^6$ & 2500 \\
7 & 10 & 35 & 4 $\times 10^6$ & 625 \\
\hline
\end{tabular}
\caption{Table listing, for each simulation, the surface density, photodissociation length scale ($l_{\rm ph}$), the number of particles and the particle mass (M$_{\rm part}$).}
\end{table}
Details of the different calculations performed are shown in Table~1. We adopt 3 different surface densities, since the degree of collisional cooling in the disc is expected to depend predominantly on the density. As described in Section~2.3.1, three different length scales are used to calculate the photodissociation rate, which determines the molecular gas abundance. The next sections describe the structure of the disc for the different surface densities, and the resulting thermal distributions. The evolution of H$_2$ is discussed in Section 3.2, including the amount of gas in the disc which becomes molecular and the timescales for H$_2$ formation. For the simulations in the main part of the paper, we use 8 million particles, but we also performed simulations with 1 and 4 million particles to investigate the dependence of our results on resolution (see Appendix).
\subsection{Structure of the disc}
\begin{figure}
\centerline{
\includegraphics[bb=40 0 560 700,scale=0.32,angle=270]{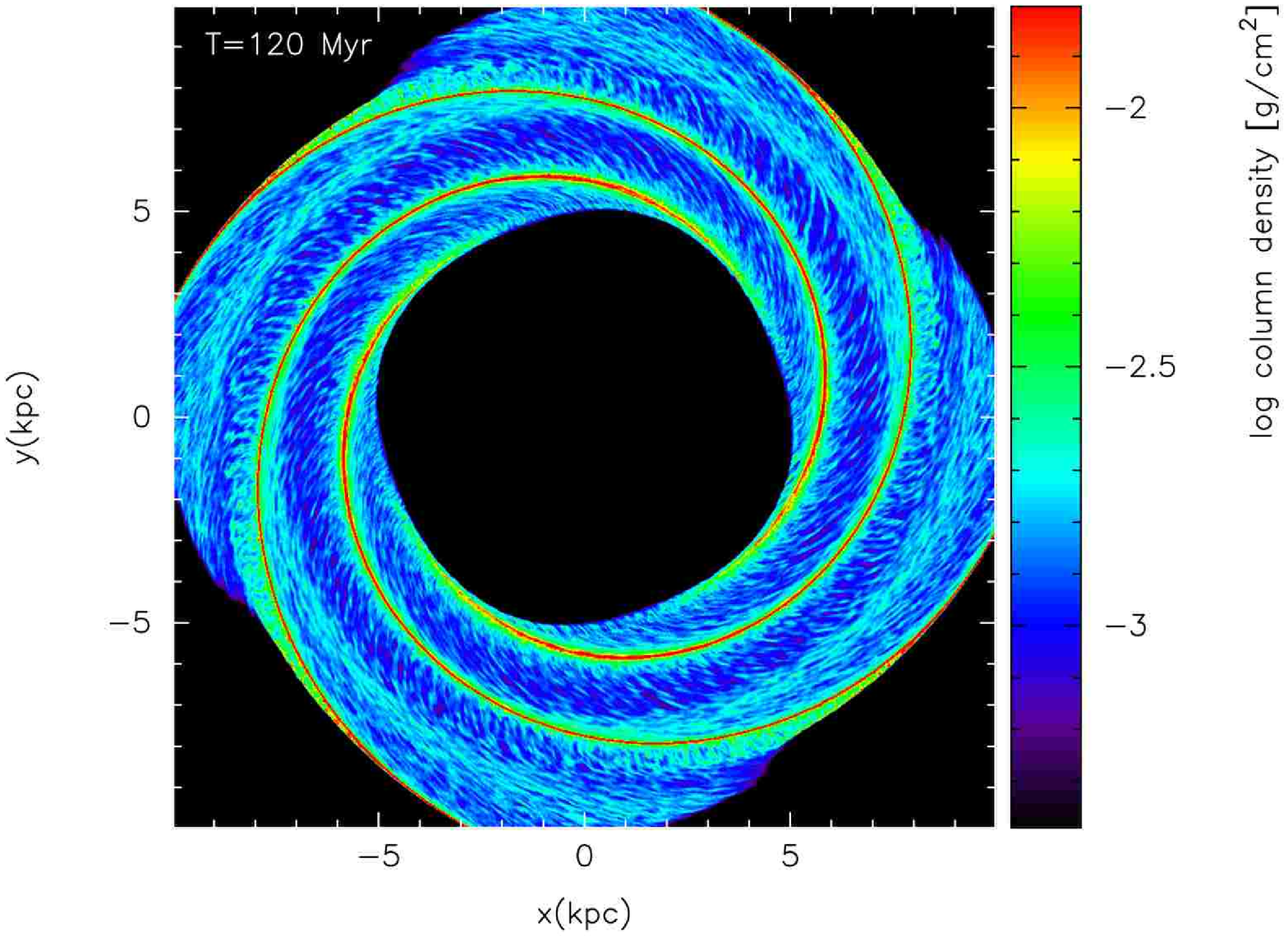}}
\centerline{
\includegraphics[bb=40 0 560 700,scale=0.32,angle=270]{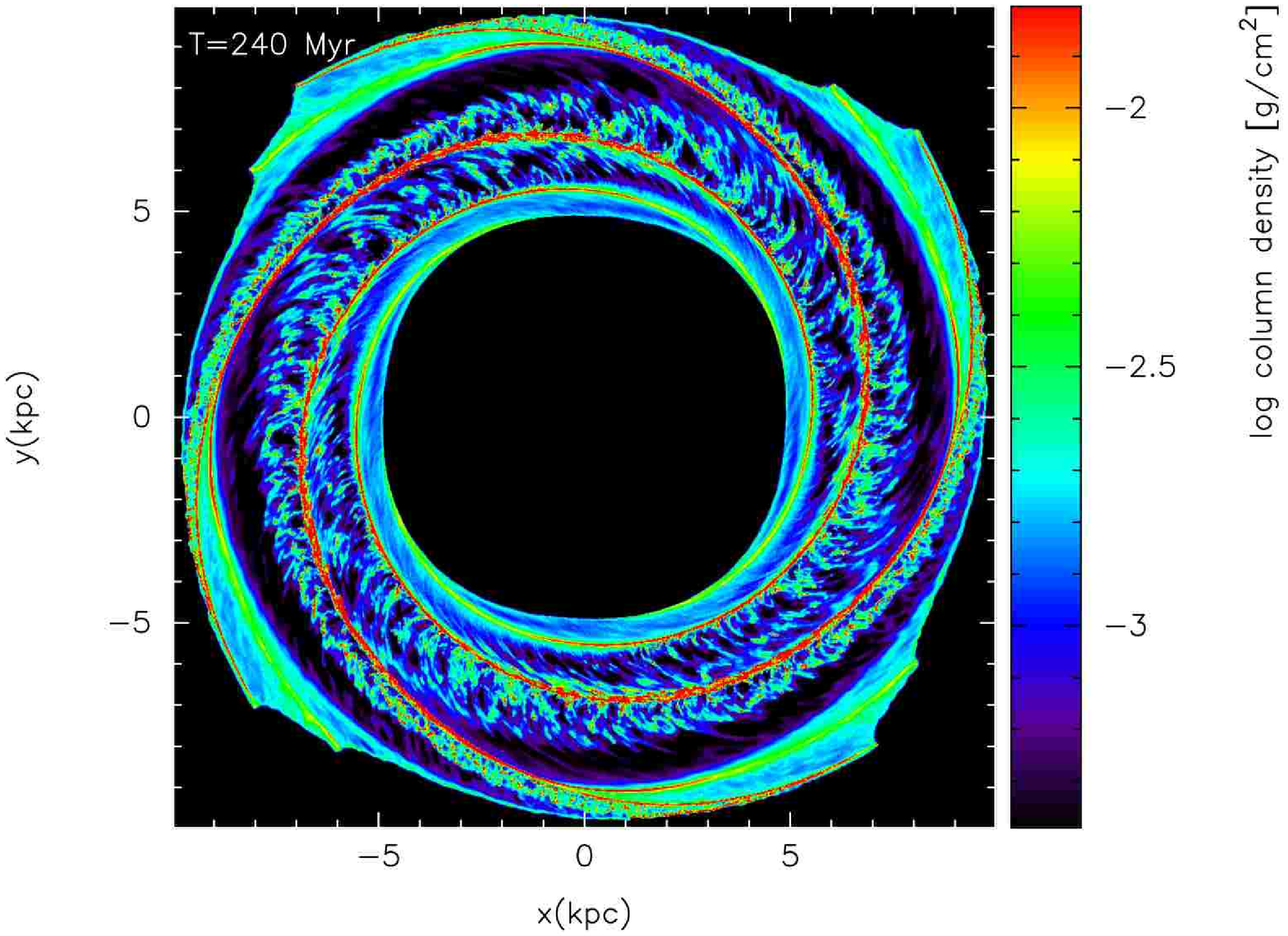}}
\centerline{
\includegraphics[bb=40 0 560 700,scale=0.32,angle=270]{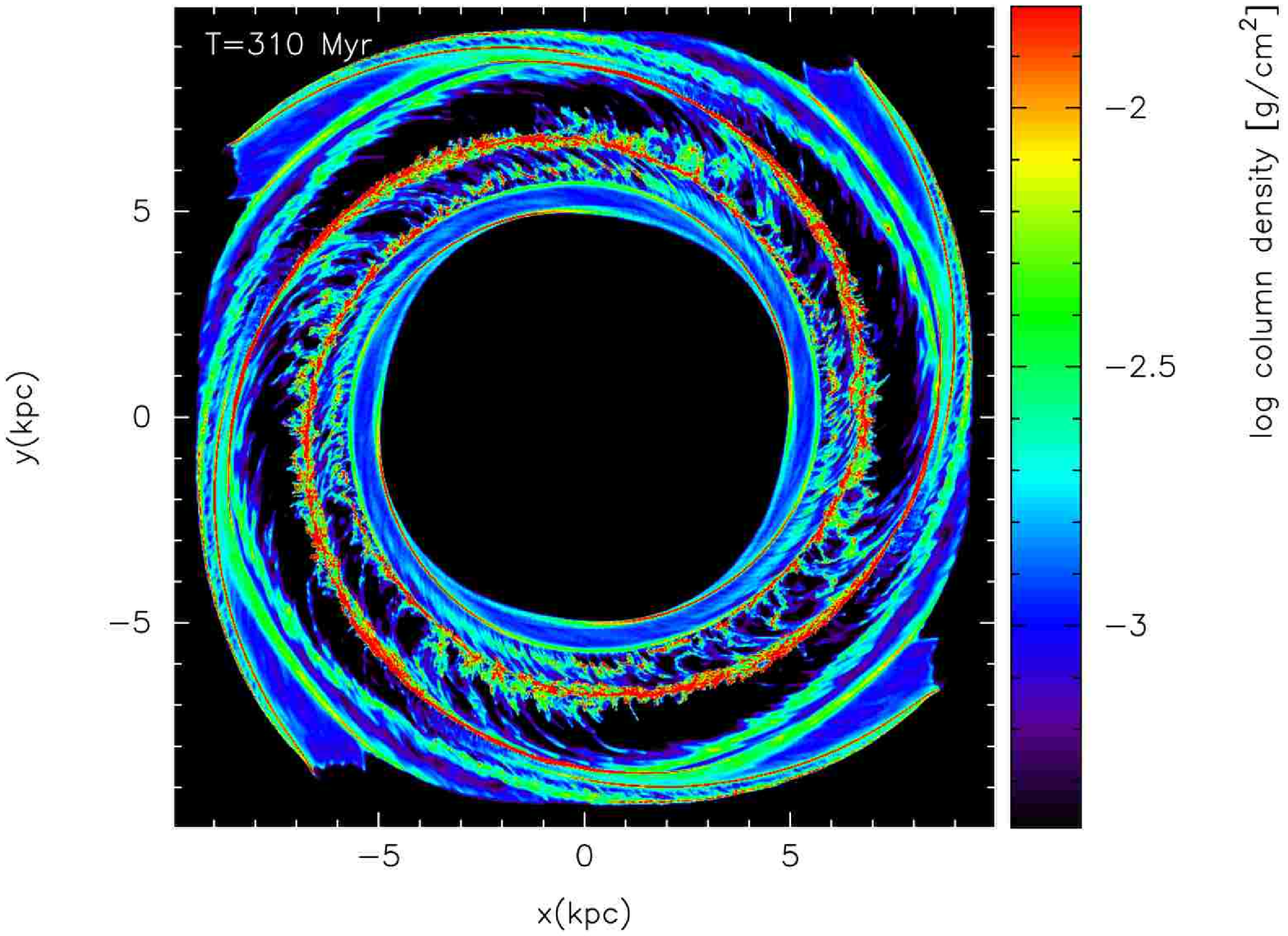}}
\caption{The evolution of the galactic disc is shown for the standard case, where $\Sigma=10$ \msp and $l_{\rm ph}=35$ pc. The disc contains a substantial degree of substructure, in agreement with previous isothermal calculations \citep{Dobbs2006}. Gas cools and becomes dense in the spiral arms, leading to the agglomeration of clumps into larger structures which shear off the spiral arms to become spurs (lower panels).}
\end{figure} 
\begin{figure}
\centerline{
\includegraphics[bb=40 0 560 700,scale=0.32,angle=270]{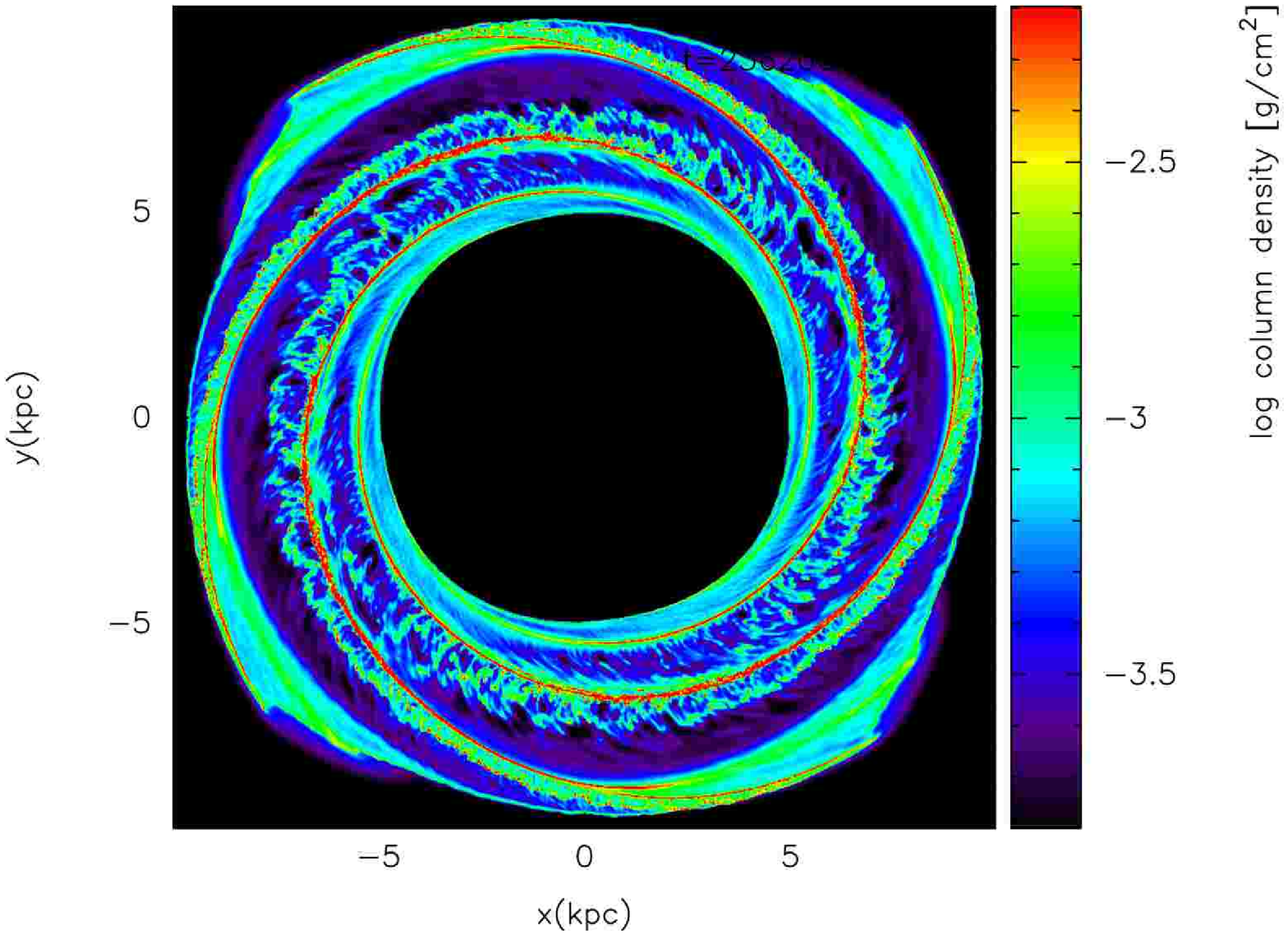}}
\centerline{
\includegraphics[bb=40 0 560 700,scale=0.32,angle=270]{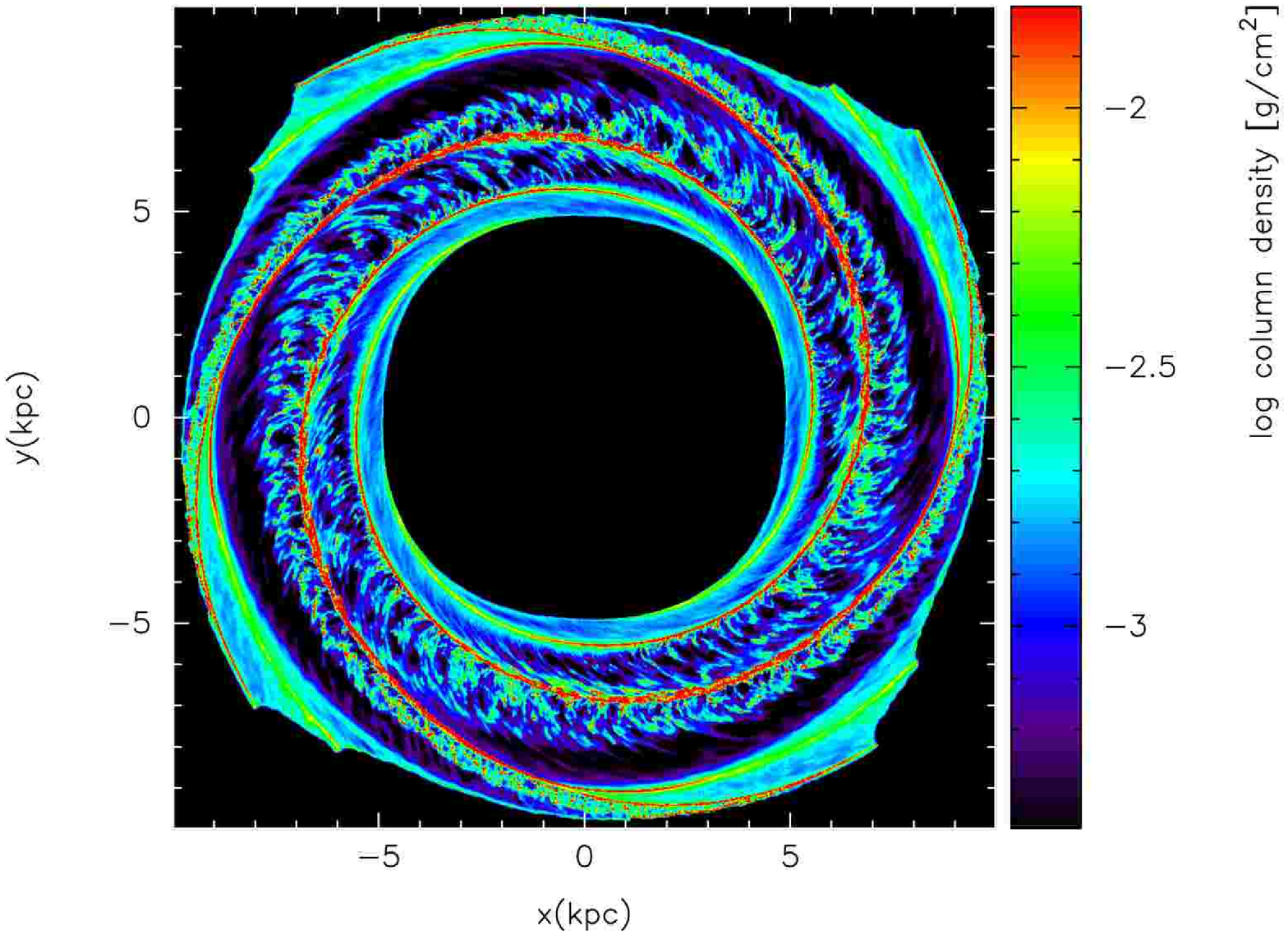}}
\centerline{
\includegraphics[bb=40 0 560 700,scale=0.32,angle=270]{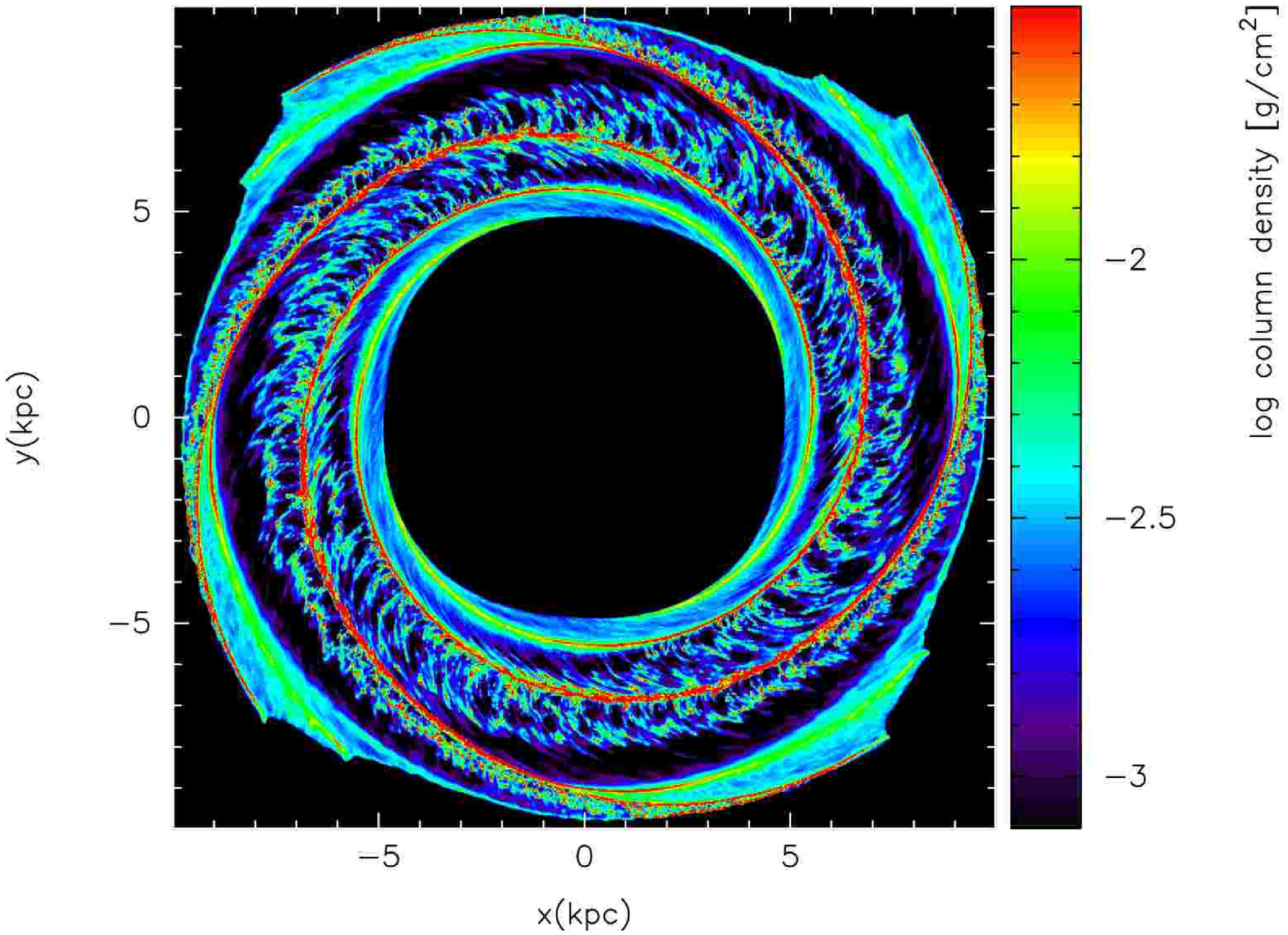}}
\caption{The column density of the disc is shown after 240 Myr for the models with surface densities of 4 (top) 10 (middle) and 20 \msp (lower). The structure is similar for all surface densities, although more high density gas is apparent in the arm and interarm regions for the 10 and 20 \msp cases. The scales have been chosen so that they are proportional to the average column density.}
\end{figure} 

The galactic disc is shown for the different surface densities in Figs~1 and 2. In all cases, the disc exhibits a large degree of substructure. The morphology of the disc is similar to that seen in previous isothermal calculations of cold gas \citep{Dobbs2006}. This is because, as will be discussed in Section 3.1.1, a substantial amount of cold gas is formed in all of these calculations, and dense features such as clumps and spurs are produced in the cold gas.

Figure~1 shows the structure of the disc for our standard model, with $\Sigma=10$ \msp and $l_{\rm ph}=35$ pc, at 3 times during the evolution of the disc. At the earlest time, there is not substantial substructure, but small spurs are beginning to emerge from the spiral arm (the transition between gas that has passed through a strong shock and gas which has not is evident: up to a radius of between 7 and 8 kpc, all the material has passed through at least one shock). At this stage, there is cold gas in the spiral arms, but little substructure has yet developed. By 240 Myr, the structure in the disc is very similar to previous isothermal calculations \citep{Dobbs2006}, with spurs extending into the interarm regions. The spiral shock causes the cold gas in the spiral arms to clump together \citep{DBP2006,Dobbs2008} and these clumps emerge from the spiral arms as distinct spurs.
 
In Fig.~2, the results for the different surface densities are displayed. The corresponding time is 240 Myr, and $\Sigma=$ 4 (top), 10 (middle) and 20 (lower) \mspnospace. Even the lower surface density case shows a high degree of substructure, since there is still a large component of cold gas. In all cases, since the cold gas is the most dense, this dominates the structure. There is little difference between the higher surface density cases. However there is comparatively more dense gas, both in the spiral arms and the interarm structures, than in the 4 \msp run. For the higher surface density discs, more of the gas reaches very low temperatures, of $\sim 10$ K. This very cold gas is denser, and remains at high densities for longer. The density is also noticeably lower in the interarm regions compared to the lower surface density case. 

\subsubsection{Thermal distribution of the gas}
\label{thermal_distrib}
In the previous section, we highlighted that cooling of gas in the spiral shocks is sufficient for gas to form dense clumps in the spiral arms, which in turn lead to interarm features. Here we discuss in more detail the thermal distribution of the gas. Overall, except for the lowest surface density case, the largest proportion of gas ends up in the cold phase.

Figure~3 shows the pressure distribution versus number density of a selection of particles from the disc. Generally there is a tight correlation between the pressure and density. The distribution corresponds to a two phase medium, with stable points at $\sim 10^4$ K and 100 K. There is however gas with $0.01 \lesssim n \lesssim 1$ cm$^{-3}$ which exhibits a lower than expected pressure. This represents gas which has recently passed through the spiral arms, but has not yet heated back up to equilibrium pressure. This gas is only a few per cent of the total amount of gas. Conversely, there are no gas particles which have higher than expected pressures, since we do not include any heating of the dense gas by supernovae.

\begin{figure}
\centerline{
\includegraphics[bb=0 0 550 380,scale=0.48]{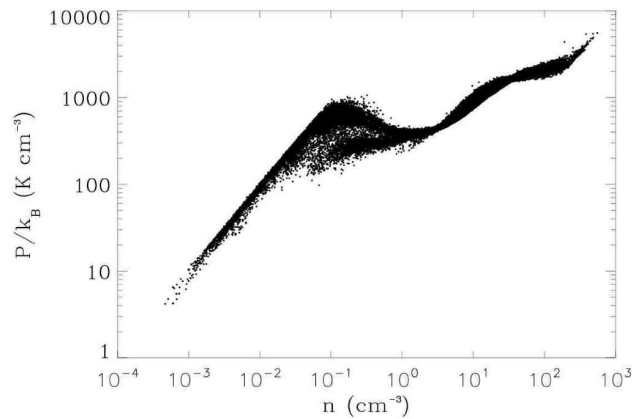}}
\caption{The pressure is plotted against number density for 100,000 random particles. The distribution shows gas at stable points corresponding to temperatures of 100 (n$\sim$20 cm$^{-3}$) and $10^4$ K (n$\sim$0.1 cm$^{-3}$) with unstable gas between these regimes.}
\end{figure} 

In Fig.~4 we show the distribution of temperature for the different surface density simulations. All show peaks at around 100 and $10^4$ K, as expected from the two phase model of \citet{Field1965}. There is more warm ($10^4$ K) gas in the lower surface density models, whilst at higher surface densities, more of the gas cools to temperatures below 100 K.
There is also a significant amount of gas in the thermally unstable regime, as has been noted in previous simulations \citep{Gazol2001,Kritsuk2002,Vaz2003,Avillez2004,Piontek2005}.
In the 10 \msp model plotted here, the photodissociation length scale is $l_{\rm ph}=35$ pc; there are only small differences in the distribution for the other length scales, mainly at low temperatures where the amount of molecular gas, and therefore the mean molecular weight $\mu$ differs. 
\begin{figure}
\centerline{
\includegraphics[bb=40 340 660 800,scale=0.43]{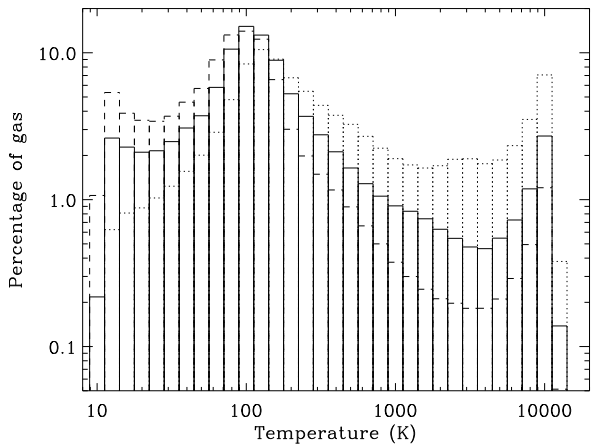}}
\caption{The distribution of temperatures for a surface density of 4 (dotted), 10 (solid) and 20 (dashed) \mspnospace. This distribution corresponds to a time of 240 Myr. Peaks are evident at 100-200 K and $10^4$ K, the stable points for cold and warm phases of the ISM respectively.}
\end{figure} 
\begin{figure}
\centerline{
\includegraphics[bb=40 360 660 800,scale=0.43]{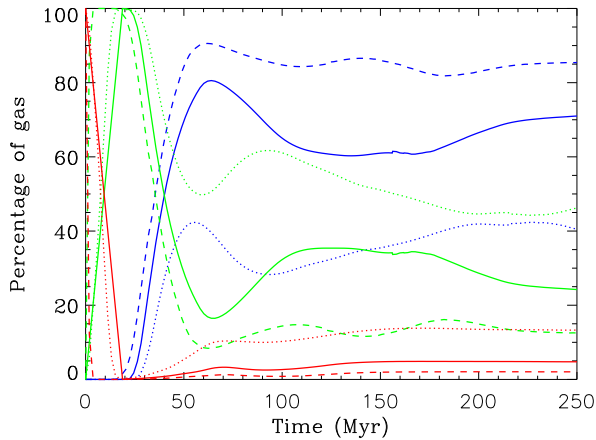}}
\caption{The percentage of gas which is cold ($T\le 150$ K) (blue), warm ($T\ge 5000$ K) (red) and in the unstable regime (150 K $\le T \le$ 5000 K) (green) is shown versus time for the 4 (dotted), 10 (solid) and  20 (dashed) \msp surface density calculations.}
\end{figure} 

\begin{table}
\centering
\begin{tabular}{c|c|c|c}
 \hline
 $\Sigma$ & 4 \msp & 10 \msp & 20 \msp \\
  \hline
 Cold & 40\% & 70\% & 85\% \\ 
 Unstable & 46\% & 25\% & 13\% \\
 Warm & 14\% & 5\% &  2\% \\
\hline
\end{tabular}
\caption{The amount of gas in the cold, unstable and warm phases is shown after 250 Myr for the three different surface density simulations.}
\end{table}

The evolution of the amount of gas in the cold ($T\le 150$~K), unstable (150~K~$\le~T~\le$~5000~K) and warm ($T\ge~5000$~K) phases is shown in Fig.~5. Clearly the amount of cold gas in the disc increases with surface density. At the outset, all the gas is in the warm regime, at 7000~K. Gas which does not immediately enter a spiral arm cools to a few 1000~K, in the thermally unstable regime. Gas entering the spiral shock cools to much lower temperatures of a few 100~K. As the shock develops further, the gas continues to cool, to temperatures of ${\sim} 100$~K.  From Fig.~5, we see that the amount of cold gas grows over a period of approximately 30 Myr, comparable to the typical crossing time in the spiral arms.  Much of the gas cools to temperatures of $<150$~K, but a small proportion of the gas also heats back up to the warm stable regime. After further spiral arm passages the amount of gas in each phase does not change significantly.

Over the course of each simulation, the evolution of the cold and unstable phases appear synchronised, whilst the amount of warm gas is fairly constant after 100 Myr. This suggests that transitions occur predominantly from the cold to unstable phases and vice versa. However, we should again stress that we do not include heating from supernovae, or from other forms of stellar feedback (e.g.\ winds) in these calculations.

By 200 Myr (approximately 2 rotations), the amount of gas in each phase is relatively stable. 
Except for the low surface density case, the largest proportion of the gas is cold. Table~2 compares the amount of gas in each phase for the different simulations after 250 Myr. There are only few observational estimates of the fractions of gas in the different regimes. \citet{Heiles2003} estimate that $\sim$ 40\% of the HI in the solar neighbourhood is cold, 30\% is unstable and 30\% is warm. In our simulations, however, the cold component includes molecular gas. Thus discounting the molecular gas (which is predominantly cold), the percentages of HI in the cold, unstable and warm regimes are approximately 62, 31 and 6\% for the 10 \msp calculation.

Several numerical simulations consider cooling and heating in local regions of the ISM, usually applying periodic boundary conditions. \citet{Piontek2005} obtain similar proportions to the 10 and 20 \msp results (which represent a comparable average $n$) for 3D simulations of MRI driven turbulence, as do \citet{Kim2008} for 1D calculations of gas subject to spiral shocks. For 2D simulations of driven turbulence, \citet{Gazol2001} find half the gas is unstable and a quarter warm or cold. \citet{Audit2005} also model turbulence and obtain relatively low fractions of cold and unstable gas ($\sim 30$\% total) but this may well be due to their boundary conditions; they use inflow conditions with a constant injection of warm gas, rather than a fixed mass of gas.

Few global simulations have investigated the thermodynamics of the ISM. \citet{Wada2001} include stellar feedback as well as the thermodynamics in their 2D models of galactic discs, although the structure formed in their simulations is due to self gravity rather than spiral shocks. They find approximately 2/3 of the gas lies in the cold regime, 20\% is unstable and the remainder is hot gas. 
Recent non-isothermal 3D simulations of flocculent spirals include \citet{Tasker2008} and \citet{Robertson2008}, but they adopt minimum temperatures of 300 and 100 K respectively, and so cannot properly address this question.

\subsubsection{Spatial distribution of warm and cold gas}
Figure~6 displays cross sections of the temperature for a section of disc (with $z=0$), for the 3 different surface densities. Evidently there is much more cold gas at higher surface densities (lower panels), which is situated in the spiral arms and the interarm clumps. The warm gas is only visible between the cold clumps, although it is fairly ubiquitous across the disc. Essentially the warm gas behaves independently of the cold, and from particle plots, the warm phase appears to shock earlier than the cold gas, as seen in Fig.~2 of \citet{Dobbs2007}.

The scale height of the warm gas is much larger than the cold (Fig.~7), with warm gas extending to 400 pc or so above the plane of the disc. The cold gas is confined within 100 pc of the plane. It is also apparent from Figs~6 and 7 that the volume of cold gas (especially below 100 K) is much smaller compared to the percentage of cold gas by mass.  
\begin{figure}
\centerline{
\includegraphics[bb=40 0 560 700,scale=0.32,angle=270]{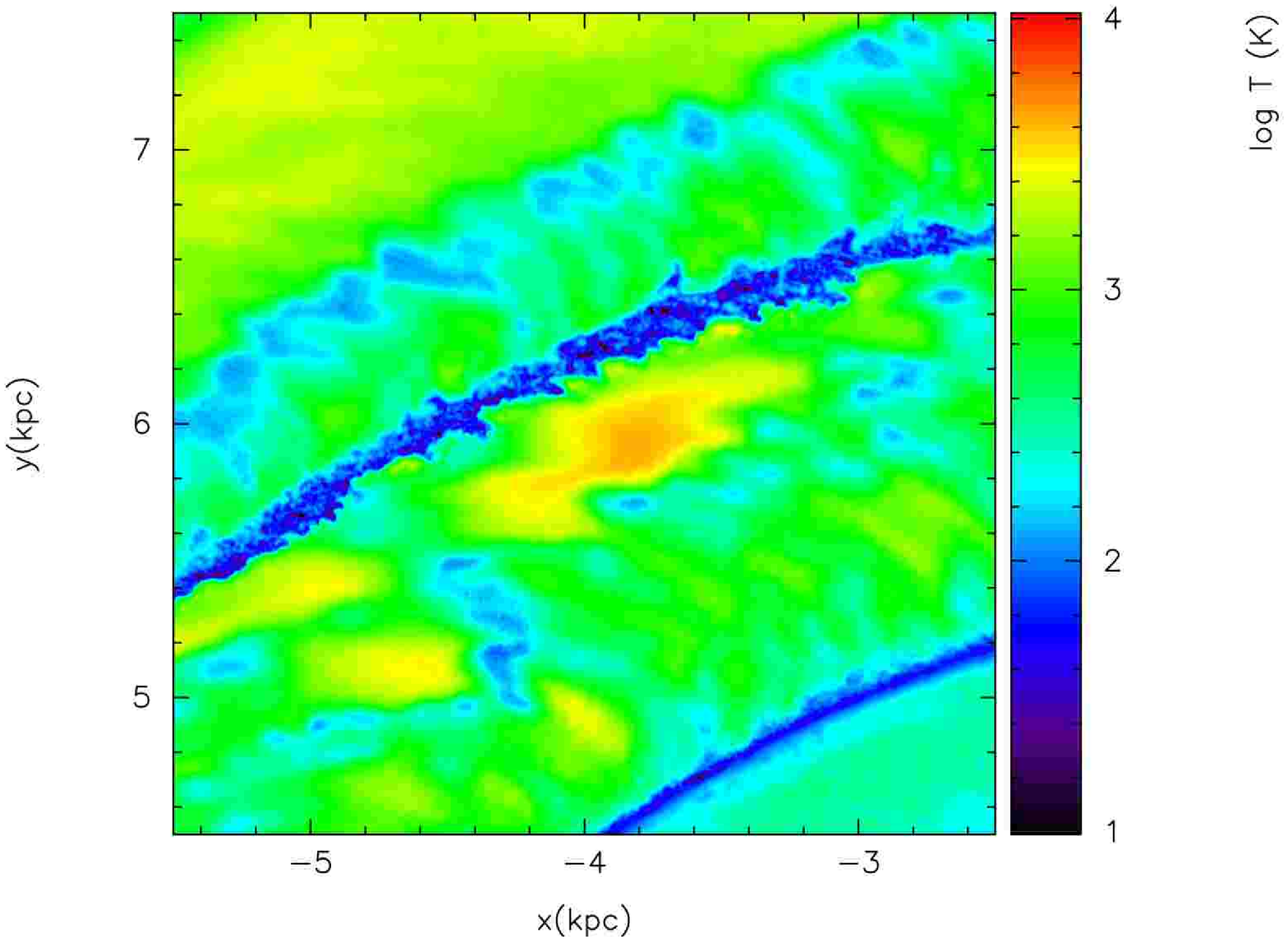}}
\centerline{
\includegraphics[bb=40 0 560 700,scale=0.32,angle=270]{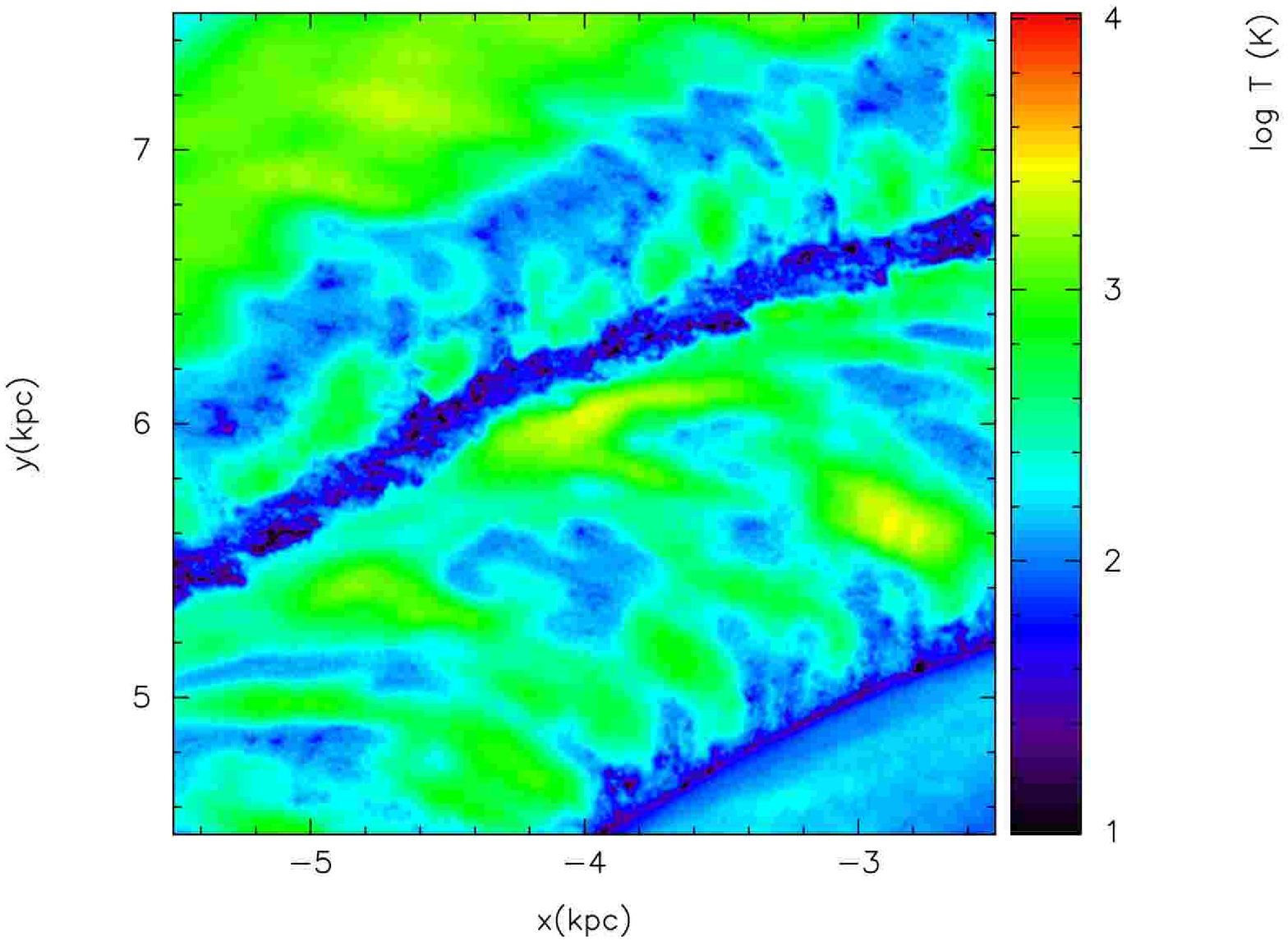}}
\centerline{
\includegraphics[bb=40 0 560 700,scale=0.32,angle=270]{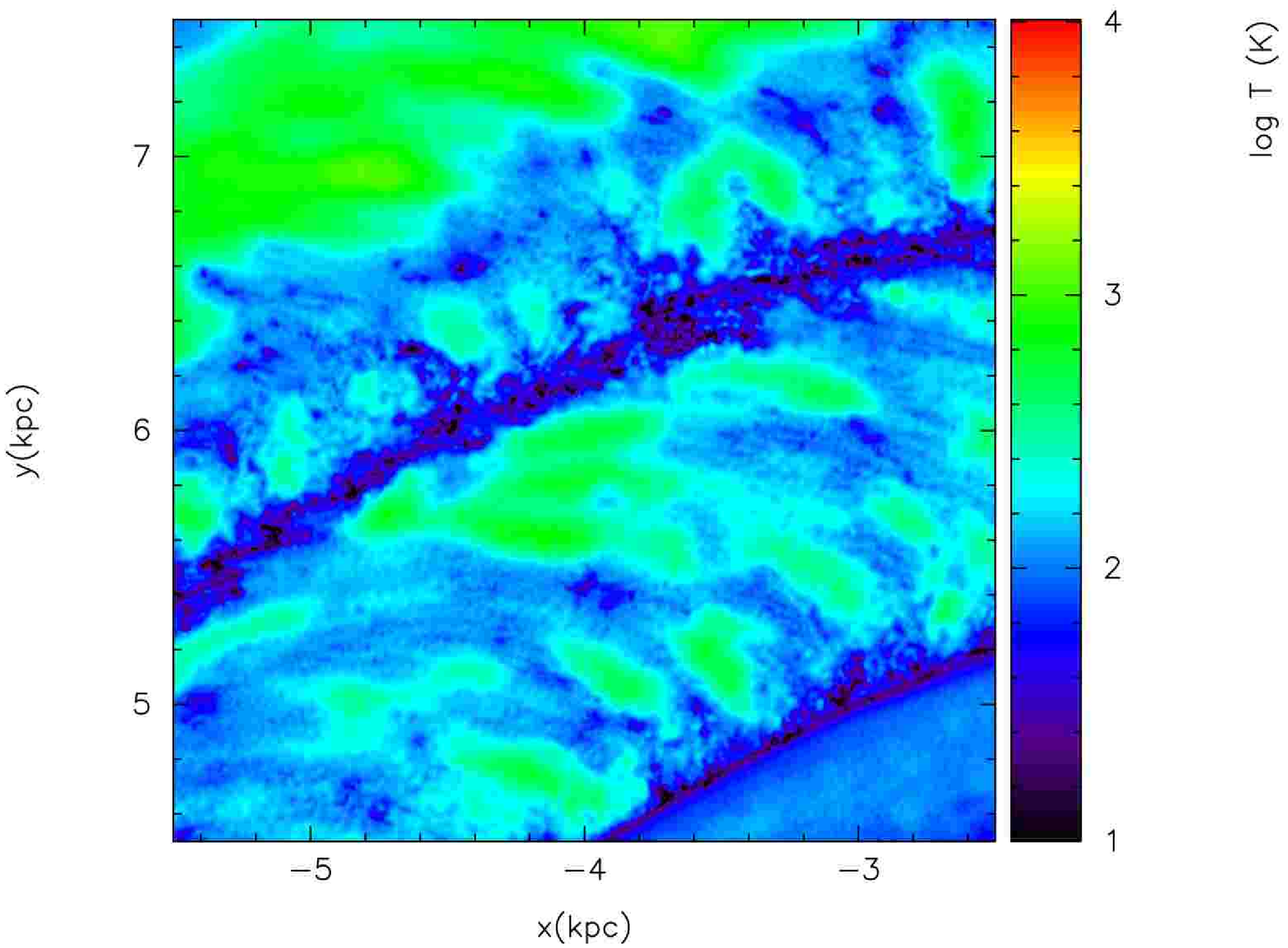}}
\caption{A cross section (where $z=0$) of the temperature is displayed for a section of the disc for the 4 (top), 10 (middle) and 20 (lower) \msp surface density discs. There is more cold gas, $<100$ K, in the higher surface density calculations.}
\end{figure} 

\begin{figure}
\centerline{
\includegraphics[scale=0.32,clip=yes,angle=270]{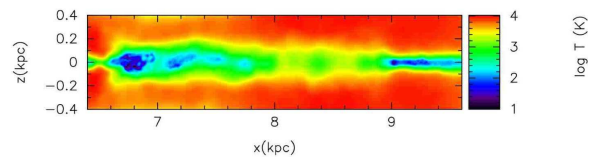}}
\caption{A cross section showing the temperature in the $zx$ plane ($y=0$) for a section of the disc. The two areas with cold gas (dark blue) on the left and right represent the spiral arms.}
\end{figure} 

\subsection{Evolution of molecular hydrogen}
The remaining part of this paper discusses the formation and evolution of molecular hydrogen. The total amount of H$_2$ in the disc is compared for the different surface densities and assumed photodissociation scale lengths in this section. The evolution of molecular hydrogen is also considered in Sections~3.2.3 and 3.2.4, in context with the other variables in the simulation, i.e. density, temperature, as well as the passage of gas through a spiral shock. We present many of the results in this section in terms of the H$_2$ fraction, which we define as $f_{\rm H_{2}}=2n({\rm H_{2}})/(n({\rm H})+2n({\rm H_{2}}))$. 

The time evolution of the percentage of H$_2$ in the disc for the different simulations is shown in Fig.~8.
As expected, the amount of H$_2$ increases with surface density, and the scale length for the photodissociation, l$_{ph}$ which determines the H$_2$ column density. From Fig.~8, the percentage of H$_2$ scales approximately linearly with the surface density. This same dependence on surface density was also found in previous isothermal calculations \citep{DBP2006}. From the lower panel comparing $l_{\rm ph}$ we see that a higher estimate of N(H$_2$) increases the self-shielding of the molecular gas (Eq.~3) and thus allows more ${\rm H_{2}}$ to survive. However, the dependence of the mean molecular fraction on $l_{\rm ph}$ is relatively weak: an increase in $l_{\rm ph}$ of almost a factor of seven increases the mean molecular fraction by no more than a factor of two. As realistic values of $l_{\rm ph}$ lie within the range 
$15 < l_{\rm ph} < 100 \: {\rm pc}$, this demonstrates that the uncertainty introduced by our simplified treatment of ${\rm H_{2}}$ self-shielding is unlikely to be very large.

The fraction of H$_2$ is seen to peak after around 180 Myr, corresponding to about 2 spiral arm passages (for gas at the midpoint of the disc). The amount of H$_2$ then levels out at approximately 7, 20 and 38 \% for the 4, 10 and 20~\msp surface density discs respectively. For comparison, at the solar radius, the surface density of H$_2$ from models of the ISM is thought to be 1 -- 2 \mspnospace, whilst the total surface density is  $\sim10~$\msp (including He) \citep{Wolfire2003}. In our $\Sigma =10$~\msp simulation, the surface density of H$_2$ is  2~\msp. Observations suggest the percentage of neutral HI which is molecular over the whole of the Milky Way is between 20 and 40\% (e.g. \citealt{Ferriere2001}).
\begin{figure}
\centerline{
\includegraphics[bb=100 300 600 780,scale=0.42]{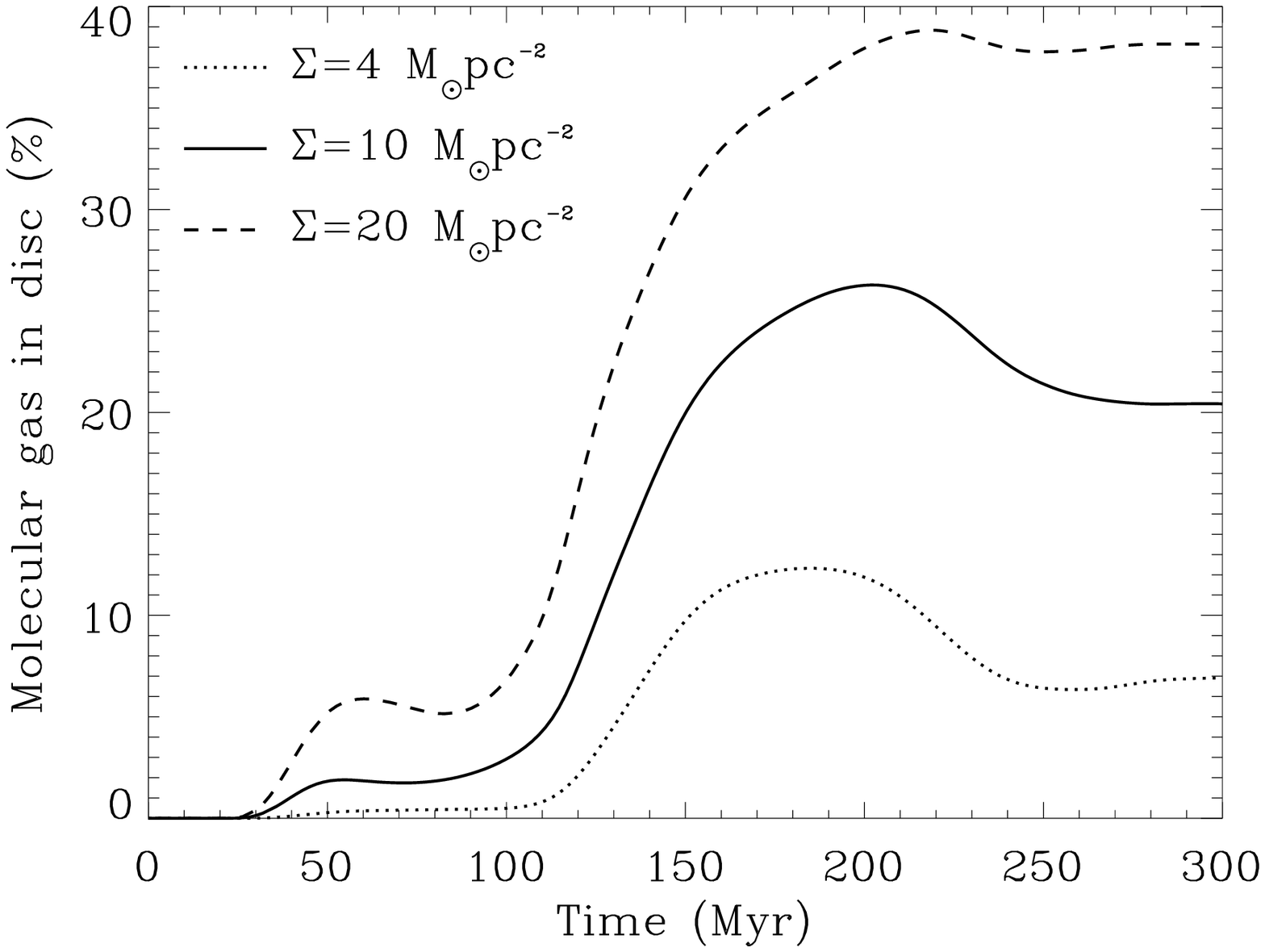}}
\centerline{
\includegraphics[bb=100 340 600 720,scale=0.42]{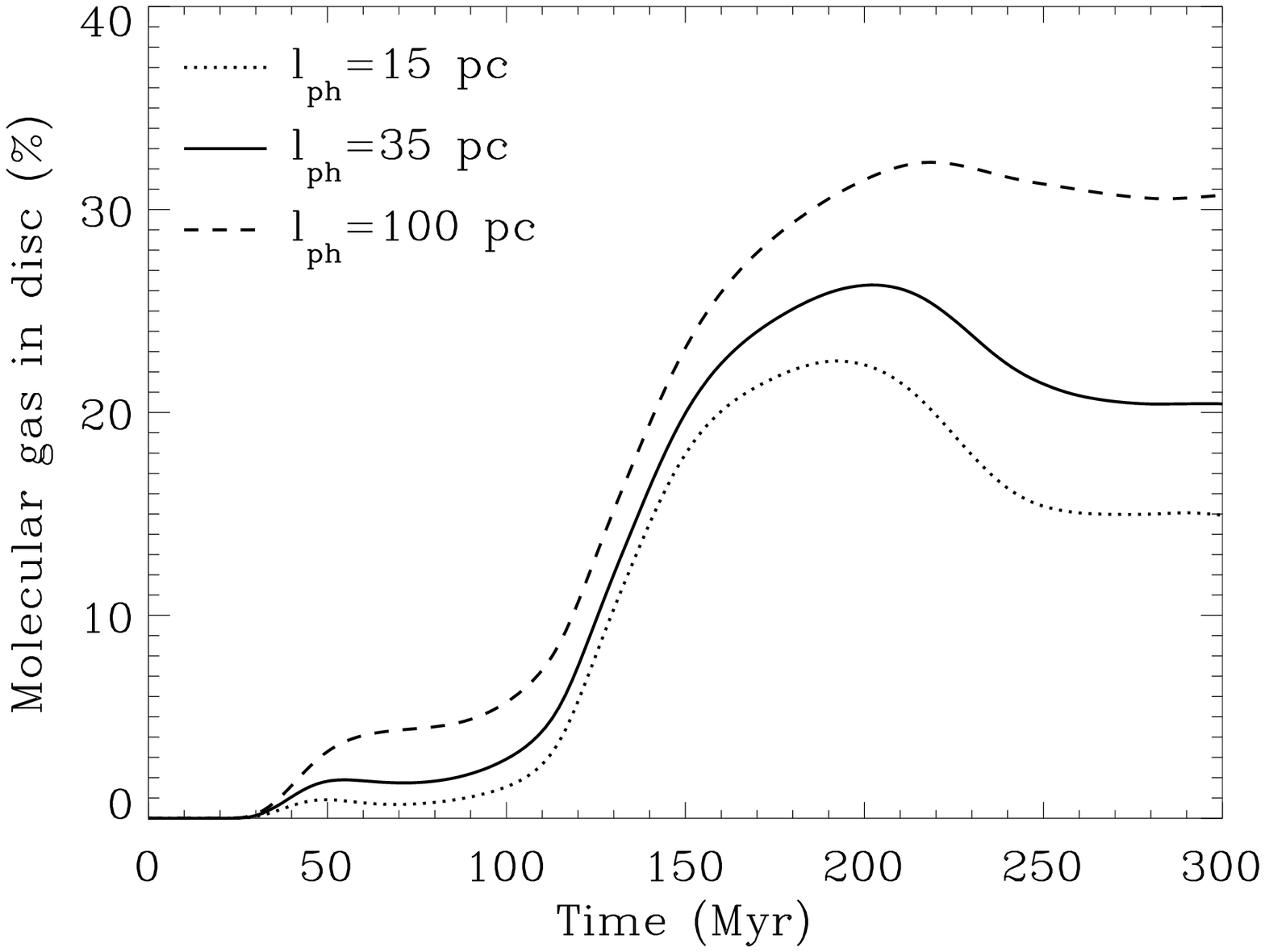}}
\caption{The percentage of molecular gas in the disc is shown versus time. In the top panel, the results for the different surface density cases are plotted, whilst the lower panel shows the cases for different photodissociation length scales (the assumed distance to a B0 star).}
\end{figure} 
  
\subsubsection{Azimuthal and radial distribution of molecular hydrogen}  
Figure~9 shows the H$_2$ fraction versus azimuth, for the fiducial case with a surface density of 10 \msp and $l_{\rm ph}=35$ pc. The H$_2$ fraction is calculated by selecting a ring located at 7.5 kpc of width 200 pc. The ring is divided into 200 segments azimuthally and the H$_2$ fraction is averaged in each. At the earlier time of 120 Myr, the H$_2$ is clearly concentrated in the spiral arms. This gas has not reached sufficiently high densities that molecular gas survives into the interarm regions. The gas heats up and almost immediately dissociates when leaving the arm. However, after 240 Myr, there are peaks situated in the interarm regions, in addition to those corresponding to the spiral arms.

In Fig.~10 we show the radial distribution of H$_2$, HI and total HI+H$_2$ after 240 Myr. The surface density is calculated by dividing the mass of gas in a ring at a given radius, by the area occupied by the ring. The HI is fairly flat across the disc, whilst the H$_2$ falls off with radius. The density of the gas tends to be underestimated at the edge of the disc though, so there is less molecular gas than expected. The decreasing H$_2$ surface density occurs since the spiral arms are further apart at larger radii and the molecular gas is predominantly in, or formed in, the spiral arms.  
\begin{figure}
\centerline{
\includegraphics[bb=90 360 600 780,scale=0.42]{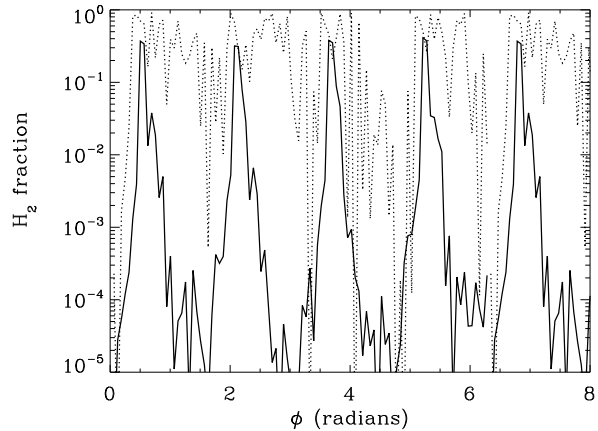}}
\caption{The molecular gas fraction is shown versus azimuth for a ring located at the midpoint of the disc, at times of 120 Myr (solid) and 240 Myr (dotted). At the earlier time, there is cold, dense molecular gas in the spiral arms, but this gas quickly warms up and photodissociates on leaving the spiral arms. On the other hand, by 240 Myr, gas in the spiral arms has reached much higher densities of H$_2$ and large clumps break away into the interarm regions, which have a high molecular gas content.}
\end{figure}   

\begin{figure}
\centerline{
\includegraphics[bb=110 380 600 800,scale=0.42]{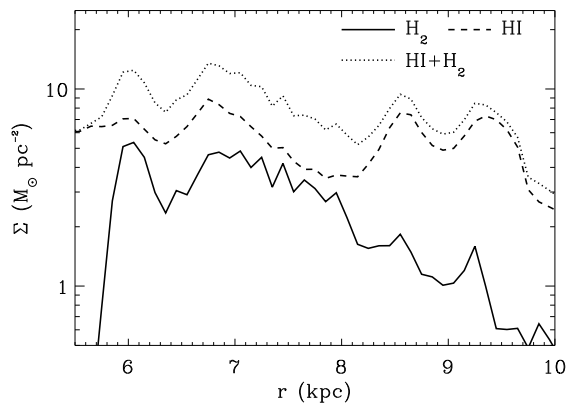}}
\caption{The surface density of HI, H$_2$ and total HI+H$_2$ is plotted versus radius for the fiducial calculation with $\Sigma=10$ \msp and l$_{ph}$=35 pc. The corresponding time is 240 Myr. The surface density profile of HI is relatively flat, whereas that for the molecular gas falls off with radius.}
\end{figure}   

\begin{figure}
\centerline{
\includegraphics[bb=90 300 600 780,scale=0.42]{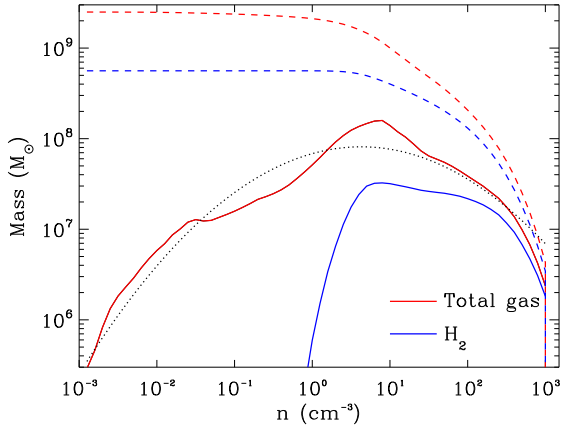}}
\caption{The mass of gas is shown versus density for the fiducial calculation with $\Sigma=10$ \mspnospace. The mass for the total gas and the molecular gas is plotted, and dashed lines indicate the cumulative mass of gas. The mass distribution for the total gas is approximated between n=10 cm$^{-3}$ and 100 cm$^{-3}$ by a lognormal distribution f($\mu,\sigma$), where $\mu=7.5$ and $\sigma=2.46$.}
\end{figure} 

\subsubsection{PDFs of the gas density}
The (mass-weighted) PDF is shown in Fig.~11, for both the total gas in the disc, and the molecular gas. The cumulative mass is also plotted. We find that the total gas PDF is approximated by a lognormal distribution, as discussed in recent simulations \citep{Wada2007,Tasker2008}. However the molecular gas does not follow a lognormal distribution. The figure indicates that there is negligible H$_2$ below $~5$ cm$^{-3}$ (the self shielding limit, see Section 3.2.3) and thus the total amount of H$_2$ above this density is the same as the total H$_2$ in the disc. 

\subsubsection{Self-shielding and the formation of molecular hydrogen} 
Whether gas in the ISM is primarily atomic or molecular is expected to depend on the efficiency of self shielding (e.g. \citealt{Federman1979,Elmegreen1993,Browning2003}).
We plot the molecular gas fraction versus density in Fig.~12. The most striking feature of these results is the large increase in molecular gas fraction at a specific density. For the fiducial case with $l_{\rm ph}=35$ pc (top panel), this density is $\sim5$ cm$^{-3}$. This is the density at which self shielding of the H$_2$ becomes effective, and the fraction of molecular gas can continuously increase with time. Below this density, the H$_2$ fraction of the gas is in equilibrium. The time to obtain molecular fractions of 0.1-1\% is very rapid, comparable to the timestep of our calculations (0.5 Myr). Hence there are few particles with molecular fractions between $10^{-6}$ and $10^{-3}$. The colour scheme indicates the mass of gas at a particular H$_2$ fraction for a given density. As will be shown later, the particles exhibit various trajectories on this plot. Gas which increases in density sufficiently to become significantly molecular follows a very similar path from low to high H$_2$ fractions, hence the large mass of gas between $n=0.1$ and 100 cm$^{-3}$.
\begin{figure}
\centerline{
\includegraphics[scale=0.42]{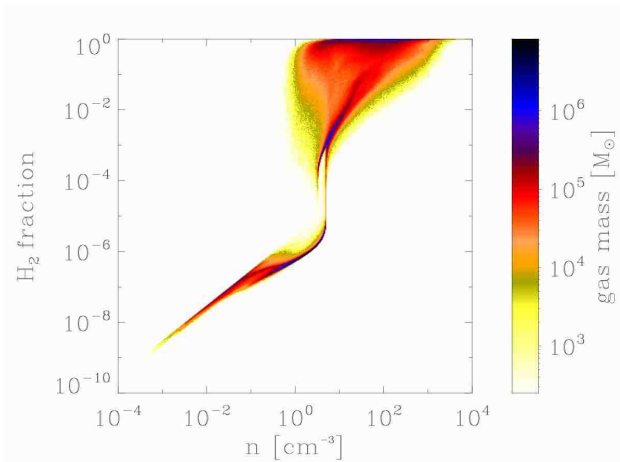}}
\caption{The fraction of molecular gas is plotted against number density for all the gas in the disc. The colour scheme indicates the amount of mass with a particular H$_2$ fraction for a given density.}
\end{figure} 

\begin{figure}
\centerline{
\includegraphics[bb=90 300 600 780,scale=0.42]{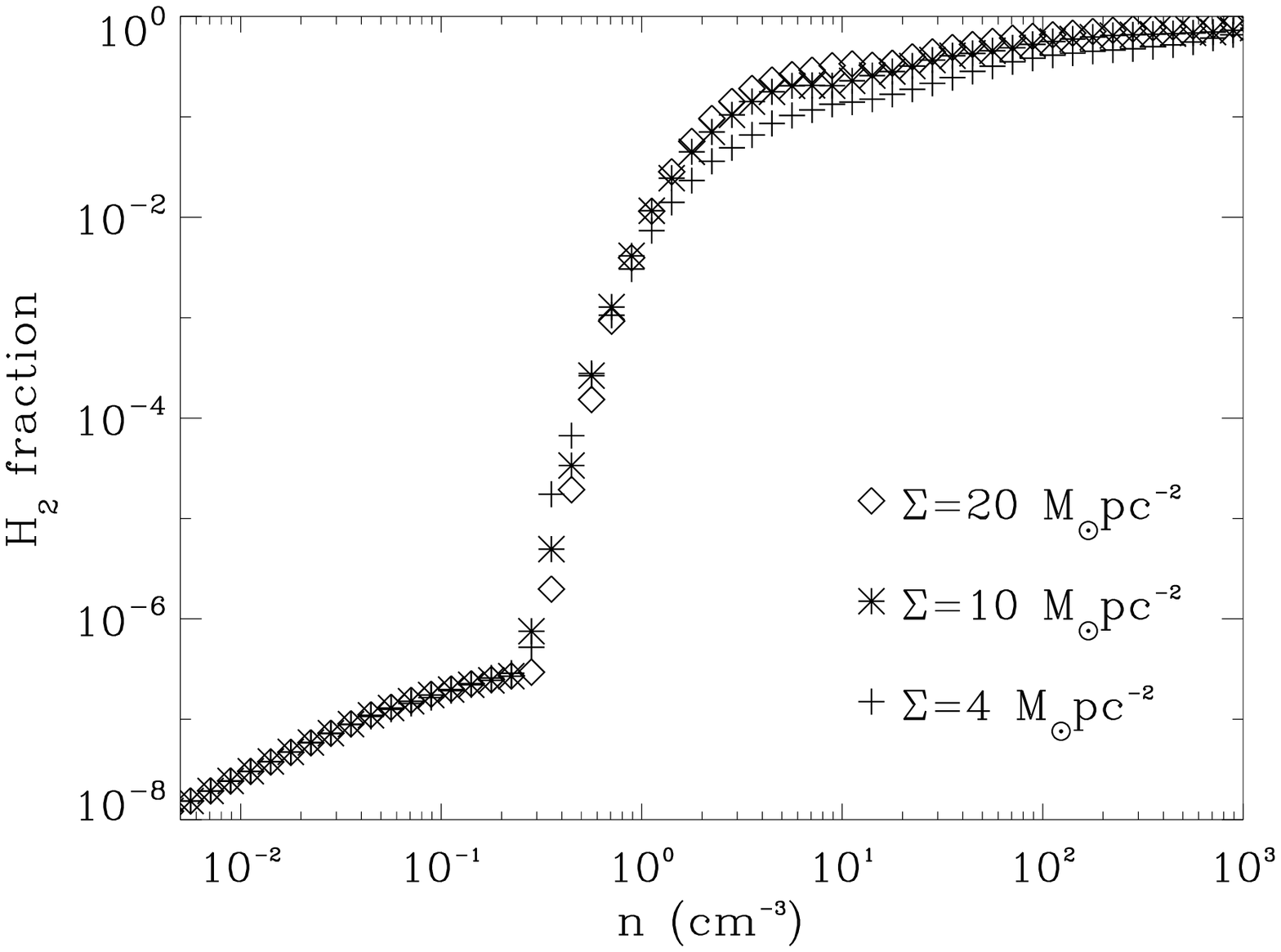}}
\centerline{
\includegraphics[bb=90 340 600 720,scale=0.42]{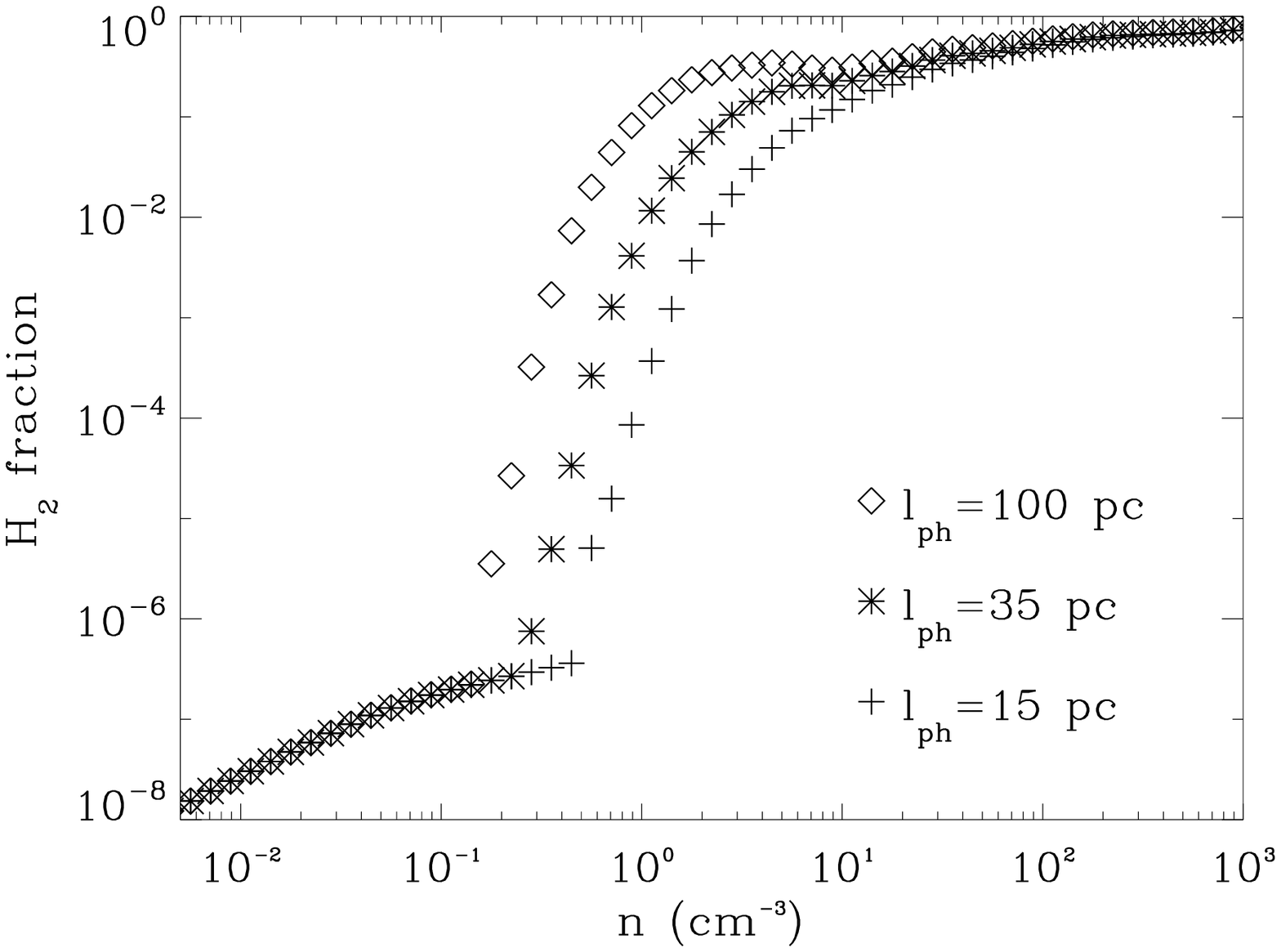}}
\caption{The molecular gas fraction is shown averaged over particles at a given number density. In the top panel, results are plotted from the different surface density calculations. The H$_2$ fraction at different densities is similar regardless of the surface density. The lower panels compare results with different photodissociation approximations. When assuming a large scale length and therefore column density, the threshold for self shielding shifts to lower densities (100~pc case).}
\end{figure} 
Figure 13 also compares the H$_2$ fraction for the different simulations. Here the molecular gas fraction is averaged over all particles at a given density, although the spread in H$_2$ fraction is similar to Fig.~12 for each case. The top panel compares the simulations with different surface density. The dependence of the H$_2$ fraction on density does not change for the different surface density calculations. Clearly when $\Sigma=20$ \msp though, much more of the gas resides at high densities, and therefore high H$_2$ fractions. The situation changes when considering different  $l_{\rm ph}$ values (lower panel). The effect of increasing $l_{\rm ph}$ is to decrease the threshold density at which self shielding becomes important. This threshold depends on $N(H_2) = n(H_2) \medspace l_{\rm ph}$ rather than $n$. So when $l_{\rm ph}=100$ pc the gas only has to exceed densities of $n \sim 2$~cm$^{-3}$ to start attaining a high molecular gas fraction (and as seen in Fig.~8, lower panel, there is more molecular gas overall), compared to  $n \sim 8$ cm$^{-3}$ when  $l_{\rm ph}=15$ pc.  

Figure~13 indicates that a large fraction of the gas is molecular at moderate densities, e.g. 10-30\% for $n=2-10$ cm$^{-3}$ and 50\% or more once the density exceeds 100 cm$^{-3}$. We stress again that this is a mean fraction, which includes gas that is currently decreasing both its density and H$_2$ fraction, possibly having already spent many Myr at densities above the self shielding limit. Furthermore the H$_2$ fraction varies with metallicity, extinction and UV flux, all of which are assumed to be constant in our model.
For comparison, in calculations performed by \citet{Pelupessy2006}, the molecular gas fraction at $n=100$ cm$^{-3}$ is $\sim$ 0.4, 0.9 or $>0.9$,  depending on metallicity, UV flux and the formation rate of H$_2$ on grains. In \citet{Koyama2000}, substantial molecular fractions are not reached until $n=1000$ cm$^{-3}$, but this is probably a consequence of lower column densities, since they consider a localised slab of gas. Finally, in \citet{Glover2007b}, turbulent clouds with mean densities of 10 and 100 cm$^{-3}$ develop molecular gas fractions of $\sim$ 0.1 and 0.4 respectively within only a few Myr, consistent with the fractions we find here.

We can also consider the evolution of particles as their density and molecular gas fraction change. Figure~14 shows evolutionary tracks of 3 particles from the simulation with $l_{\rm ph}=35$ pc and $\Sigma =10$ \mspnospace. The top panel is the most typical (i.e. frequent) path for a particle. As a particle enters a spiral shock, its density increases, and the particle evolves along the right hand side of the distribution of particles in Fig.~12. At some point, the density of the particle starts to decrease, and the particle evolves to the left in H$_2$ fraction versus $n$ space. This accounts for the large 
spread in the H$_2$ fraction apparent for densities $>5$ cm$^{-3}$ in Fig.~12. But although the particle's density decreases, it still retains a high molecular gas fraction. Thus the evolution in H$_2$ fraction is quite different depending on whether the particle's density is increasing or decreasing. The reason for this is because once the 
fluid element has attained a high molecular gas fraction, the column density $N(H_2)$ is large and self shielding (denoted by $f_{\rm shield}$ in Eq.~2) prevents the H$_2$ from rapidly becoming photodissociated. Thus the gas can exhibit higher H$_2$ fractions at lower densities compared to before the gas enters the shock (where $N(H_2)$  and the degree of self shielding is negligible). A similar scenario is described in \citet{Pelupessy2006}, who even find that molecular gas can survive the transition from the CNM to the WNM. However in reality many of the trajectories in Fig.~12 would be interrupted at high densities by star formation processes.

The lower panels in Fig.~14 indicate the evolution of two gas particles which behave somewhat differently. The middle panel indicates an example where a particle exhibits a high molecular gas fraction for most of the calculation. This is because once the gas attains a large molecular fraction, it encounters another spiral shock before becoming photodissociated to a very low H$_2$ fraction. The gas particle in the lowest panel exhibits the opposite behaviour. 
It lies in the warm regime, at a larger scale height. The gas does not experience a shock since it is too warm and diffuse. Thus the density does not become sufficiently high for the fluid element to attain a significant H$_2$ fraction.

\subsubsection{Evolution of physical variables during the simulation}
This section describes how the density, H$_2$ fraction and temperature evolve with time for specific gas particles in the simulation. Fig.~15 shows the evolution of the various physical quantities with time, as well as the distance to the nearest potential minimum, 
for the three particles already depicted in Fig.~14. The corresponding galactic radii for the particles are 7.4 kpc (top), 5.1 kpc (middle) and 6 kpc (lower panel). For particles at larger galactic radii, there are correspondingly fewer minima in the variation of the distance to the potential minimum with time, indicating fewer passages through the stellar spiral arm. 
\begin{figure}
\centerline{
\includegraphics[bb=100 320 600 770,scale=0.4,clip=yes]{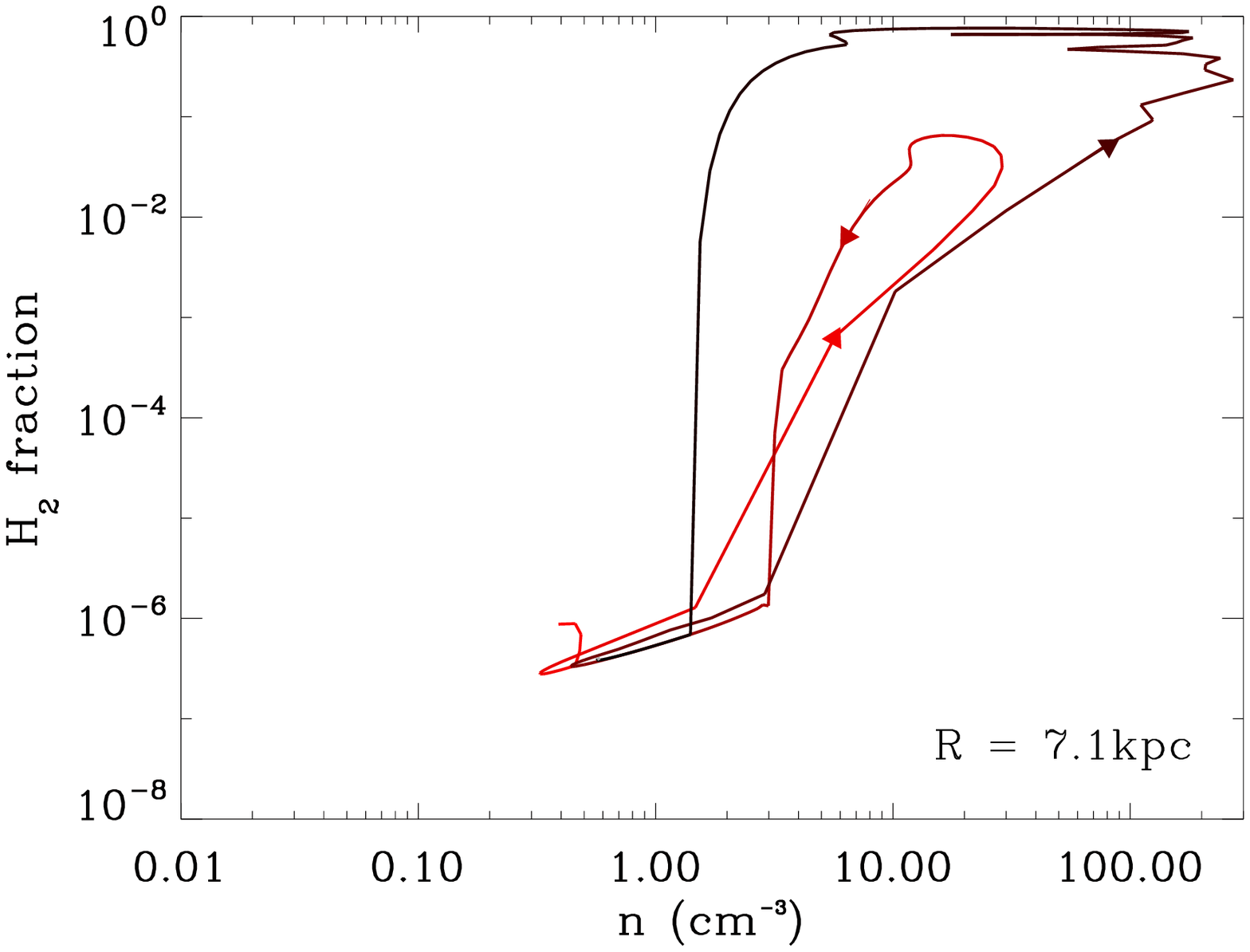}}
\centerline{
\includegraphics[bb=100 320 600 742,scale=0.4,clip=yes]{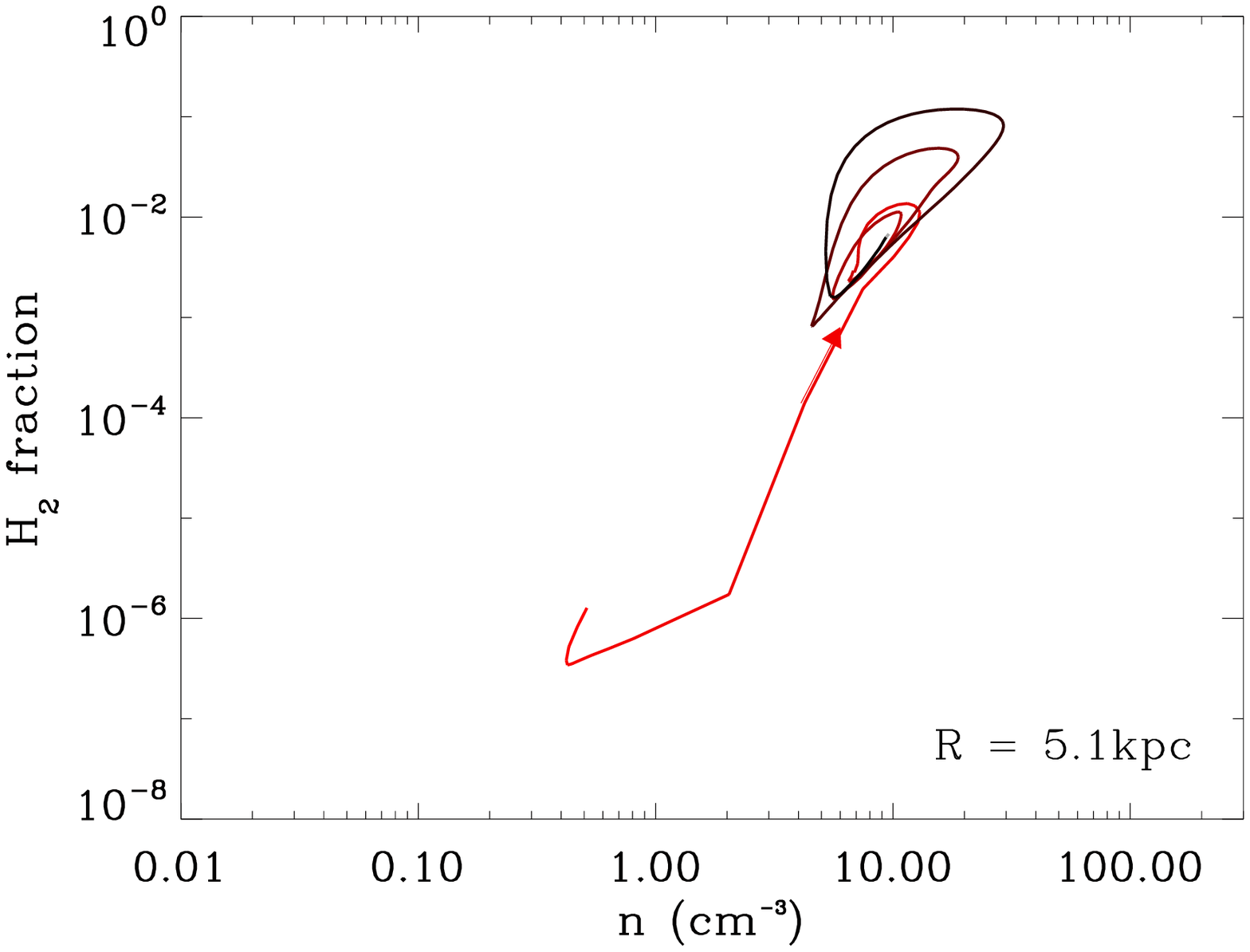}}
\centerline{
\includegraphics[bb=100 320 600 742,scale=0.4,clip=yes]{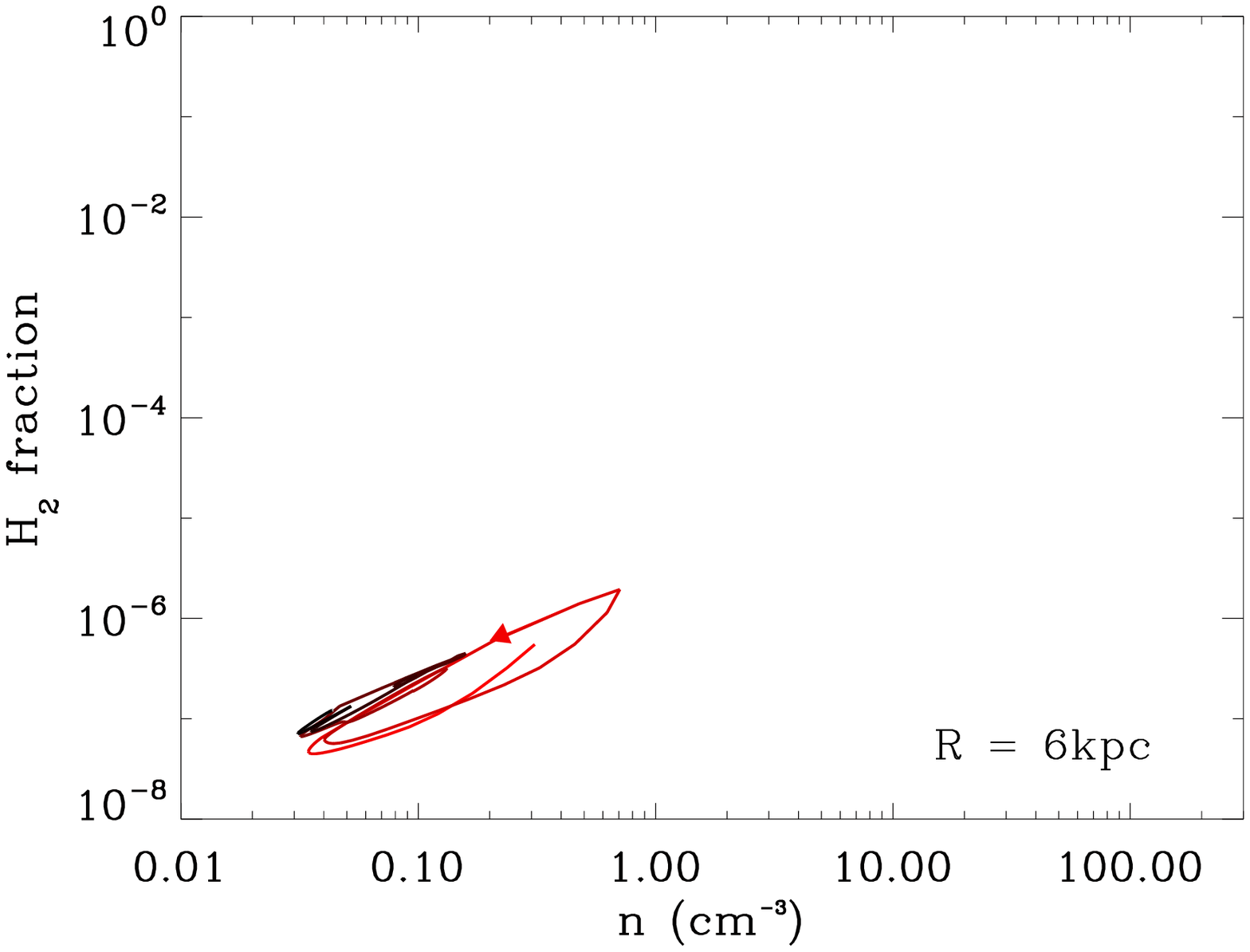}}
\caption{The trajectories of three particles are shown in H$_2$ fraction versus density space. The colour of the plotted lines changes with time, with red at t = 0 Myr to black at 310 Myr (arrows also indicate the direction the particles evolve over time). In the top panel, the gas (twice) reaches a high molecular gas fraction and then drops down to lower densities with very low molecular gas fractions. This pattern is typical for most particles in the disc. The average galactic radius for each of the particles is also indicated on the panels. The evolution of H$_2$ fraction, density and temperature are shown with time in the corresponding plots in Fig.~15.}
\end{figure} 
\begin{figure}
\centerline{
\includegraphics[bb=100 320 600 770,scale=0.4,clip=yes]{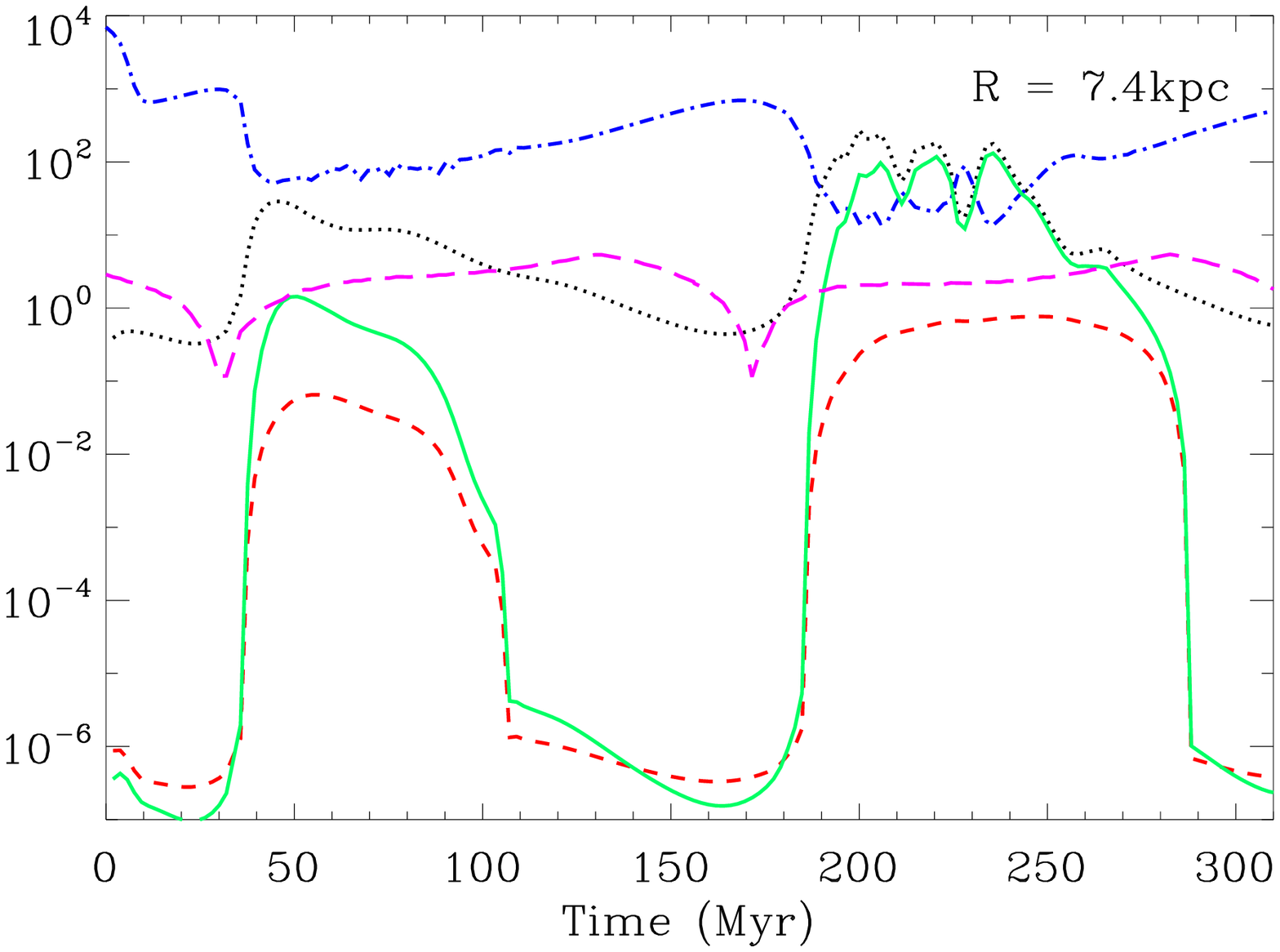}}
\centerline{
\includegraphics[bb=100 320 600 742,scale=0.4,clip=yes]{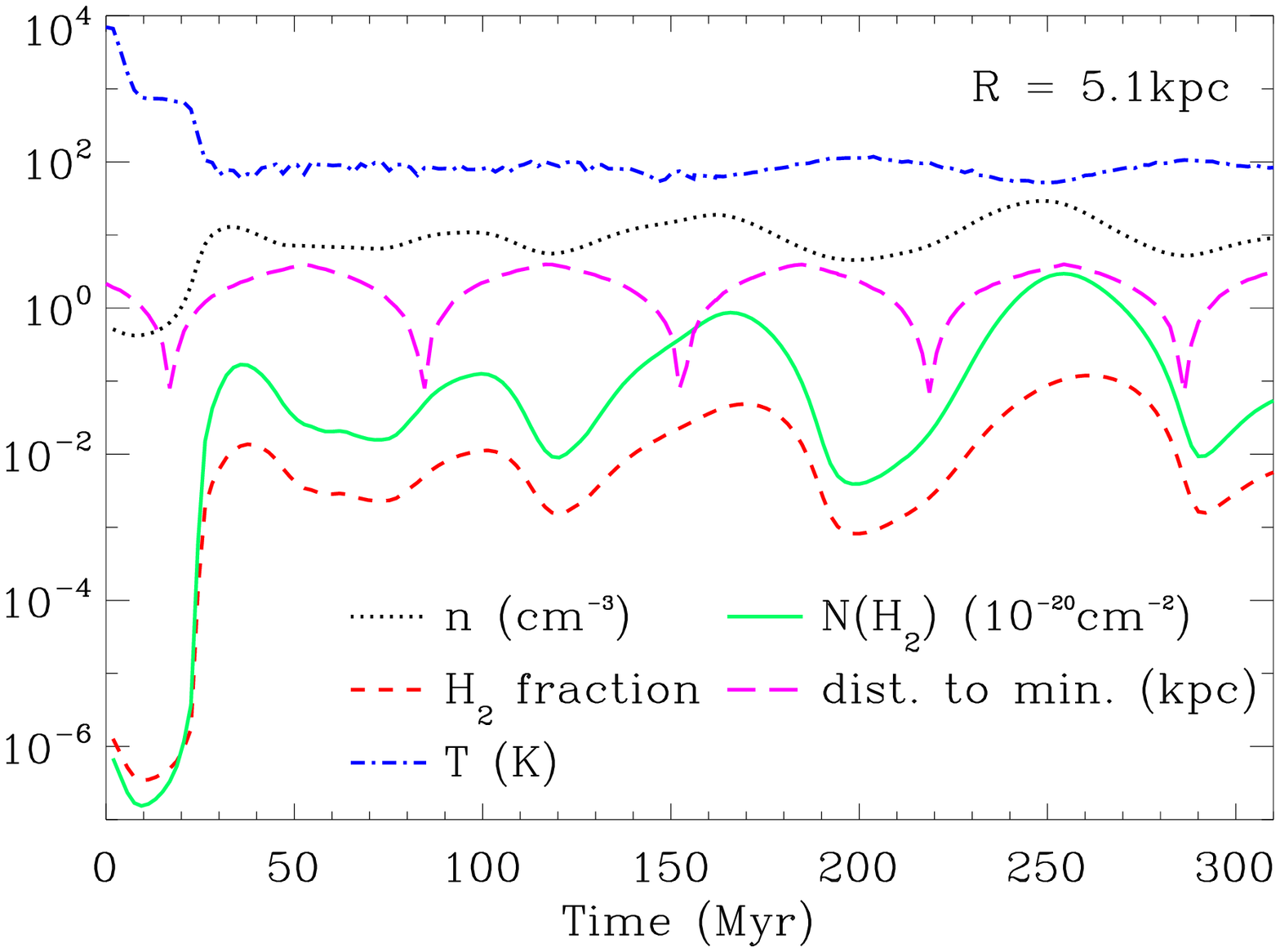}}
\centerline{
\includegraphics[bb=100 320 600 742,scale=0.4,clip=yes]{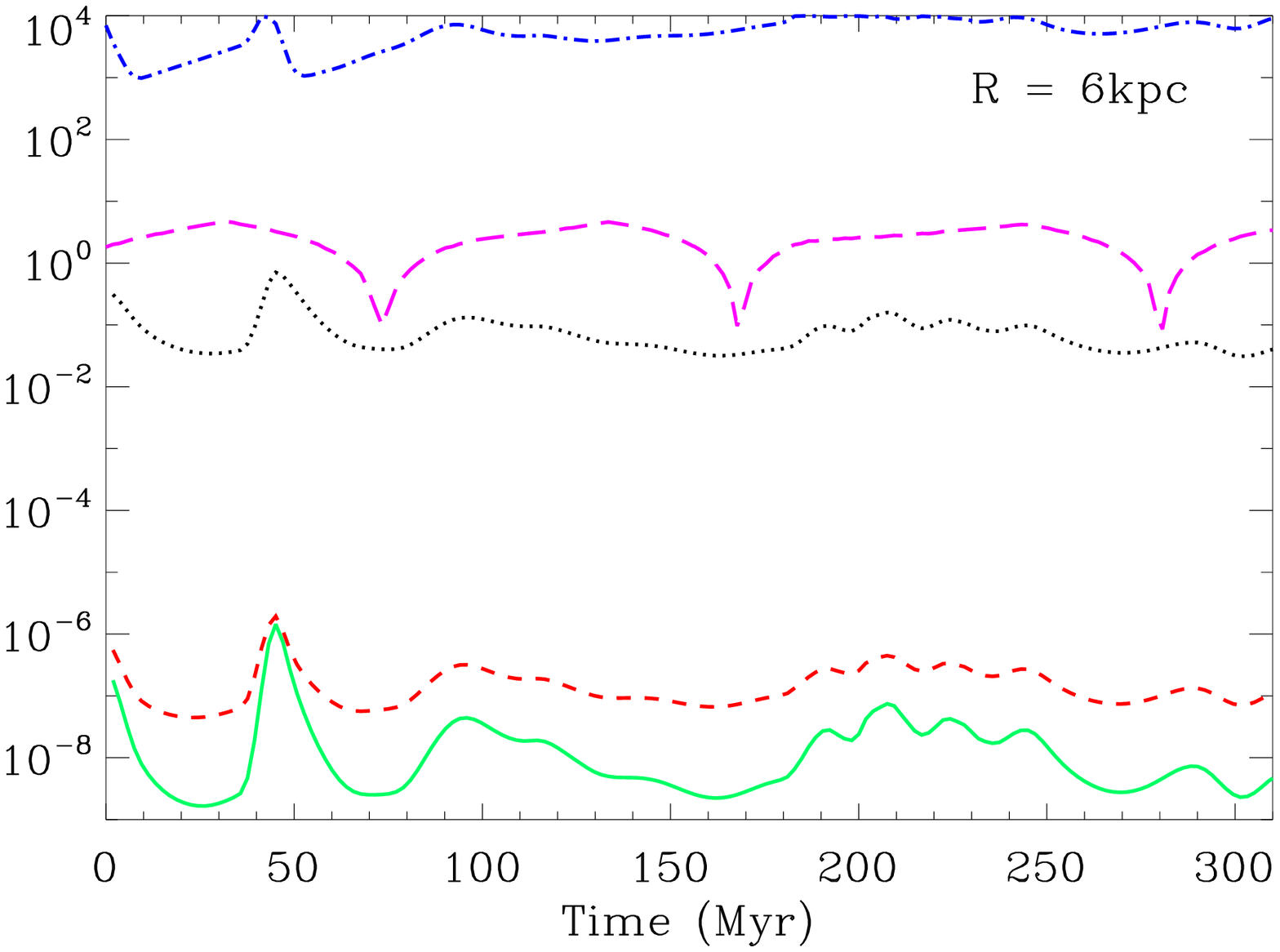}}
\caption{The evolution of density, H$_2$ fraction, temperature and column density is shown versus time for 3 particles (a different particle in the top, middle and lower panels). The distance to the nearest potential minimum was also calculated and is shown in these figures. The top panel shows fairly typical behaviour as the gas passes through the spiral shock, where the H$_2$ fraction increases in the spiral arm and then decreases. In the middle panel, after the first spiral shock, the gas retains a high density and H$_2$ fraction throughout the simulation. For the third panel, the gas density never gets very large, the gas has negligible H$_2$, and remains in the warm phase of the ISM. The galactic radii of the particles are indicated on each of the panels.}
\end{figure}

The top panel shows the typical behaviour of gas as it travels round the disc. After 30 to 40 Myr, the gas particle passes the potential minimum and enters a spiral shock. The gas experiences a sharp increase in density, and a corresponding drop in temperature to $\sim100$~K. The increase in molecular gas fraction is similarly rapid, changing from $\sim10^{-6}$ to $\sim10^{-3}$ in a single timestep (0.5 Myr). A further increase to the peak fraction of $\sim0.1$ takes around 10 Myr. The gas then remains in the spiral arm for another 40 to 50 Myr. After leaving the spiral arm, the gas becomes less dense and heats up. At the same time, the gas experiences a sudden drop in the H$_2$ fraction, decreasing to $\sim10^{-6}$ in 10 Myr or so. The gas enters a spiral shock for the second time after 190 Myr. By this time, the spiral shock is more developed and the gas reaches higher densities. Consequently 
it cools to $\sim$20~K and attains a molecular gas fraction of 0.7. Again the time to acquire this fraction is short, around 10 Myr. At this point in the simulations there is much more interarm structure, with clumps breaking off into the interarm regions. The gas retains a higher density and H$_2$ fraction for a longer period, including time when gas has left the spiral shock, but drops down to lower densities and a very small H$_2$ fraction by 300 Myr.

The middle panel in Fig.~14 shows the evolution of gas which retains a high density throughout most of the simulation (as also described in Section~3.2.3). This gas is nearer the inner edge of the disc and experiences 
more frequent spiral arm passages. The density does not decrease sufficiently for the H$_2$ fraction to return to $10^{-6}$. Consequently the molecular gas fraction stays between $10^{-3}$ and 0.1, whilst the temperature remains around 100 K (thus the gas remains in the cold regime for the rest of the simulation). Finally in the third panel, the gas never attains a significant fraction of H$_2$. The gas remains too warm and diffuse throughout the simulation.

We compared the profiles of 100 random particles, finding that the majority of particles (60-70\%) exhibit evolutionary patterns similar to the top panel of Fig.~14, i.e.\  
the molecular gas fraction increases and decreases by orders of magnitude during and after each spiral passage. A few particles  have high densities and molecular gas fractions for most of the simulation, similar to the middle panel, but these tend to be nearer the centre of the disc. A particular gas particle does not always exhibit the same behaviour through each spiral arm, e.g. gas may be low density and not molecular at the beginning of the simulation, but increase its peak density during successive spiral arm passages. Conversely a gas particle may be dense and have a high molecular gas fraction before 100 Myr, but then move into a low density region between spiral arms, and become low density atomic gas for the rest of the simulation.

\subsubsection{Timescales for the formation and destruction of H$_{2}$}
The questions of how quickly molecular clouds form, and how long they live, are
central to any model of star formation. The simulations performed for this study can
provide an insight into these timescales.

Figure 16, shows  the time taken for the gas to achieve an H$_{2}$ mass fraction of 0.5. The
characteristic time required for raising the molecular fraction from 0.1 to 0.5 is around 10 Myr, 
which is roughly the time needed to cross the spiral arm. We stress here that the  values given in
Fig.~16 are {\em upper} limits, since we do not have the resolution to treat
the internal turbulent motions in the spiral arms fully consistently. \citet{Glover2007b} demonstrated that turbulence at and below the
GMC scale can induce H$_2$ formation on an internal crossing time. \citet{Dobbs2007} showed that when
structured gas is passed through a shock it generates turbulence that is consistent with the typical
velocity scaling laws found in molecular clouds \citep{Larson1981,Heyer2004}. The fact that the typical
timescale for going between molecular fractions from 0.01 to 0.1 is faster than from fractions 0.1 to
0.5 is indicative of the resolution constraint our calculations. It is therefore likely that the
unresolved velocity structure in the spiral arms would accelerate the formation of H$_{2}$ to rates
higher than we quote here. Self-gravity would increase these rates further still and as such, one would
expect the peaks in the formation time distributions to move to lower values.

The ability of a parcel of gas (in this case, an SPH particle) to gain a large H$_{2}$
fraction, depends on its path through the spiral arm, and hence the level of compression that it experiences. Two effects are important here. The first is `orbit crowding',
whereby gas streams from different radii are brought together by the spiral potential due to their
relative motions. This increases the density of the gas that undergoes a shock during its passage
through the spiral. The second effect is simply that gas entering the spiral arm  will undergo a
stronger shock if the gas already residing in that part of the spiral  arm is cold.

At earlier stages in the simulation, the spiral shock is not well developed. Warm gas experiences compression and achieves low ($\sim$100 K) temperatures, but usually only moderate H$_2$ fractions (e.g. Fig.~9,  Fig.~15 top panel). However, subsequent gas entering the shock now encounters cold gas. This gas experiences not only compression by the shock, but collides with cold clumps in the shock. As a result, the gas experiences stronger compression, cools to lower temperatures and can obtain a higher H$_2$ fraction. With time, this process repeats, and the gas in the spiral shock becomes colder and more molecular. Eventually a steady state is reached (Fig.~8), by which time gas can enter the shock and achieve high H$_2$ fractions in 10 to 20 Myr. The formation of GMCs is thereby facilitated by the pre-existence of dense, cold material, as hypothesised by \citet{Pringle2001}.

The density reached in the spiral arms
also depends on the density and temperature of gas entering the shock (e.g Fig.~15, middle and lower panels). Gas which is in the midplane of the galaxy, or which has not fully heated up after passing the previous arm, is denser and colder and is more likely to achieve a high H$_2$ fraction (Fig.~15, middle). As discussed in  Section 3.3.1, much of the gas lingers at
densities around 0.1 to 1 cm$^{-3}$ and  temperatures of several hundred Kelvin. This places it in
the regime where self-shielding starts to become important (see Figures 11 \& 12). As such, the gas 
remains in a regime where it can easily increase its molecular fraction during the  compression produced by the spiral arm passage. Gas at low density, e.g. above the mid plane, may pass through the spiral arm without obtaining a significant increase in H$_2$ fraction (Fig.~15, lower).    
Overall, the combination of orbit crowding and collisions 
between parcels of cold gas that is described above creates large cold molecular complexes. Between passages through the spiral arms, these parcels of cold gas typically do not maintain a high H$_{2}$ fraction.

Finally Fig.~17 indicates the  
time gas spends above given molecular fractions. The gas tends to spend more time at higher H$_2$ fractions. We can relate this behaviour to Fig.~14, which shows the trajectories of particles in H$_2$ versus $n$ space. The particles' trajectories form loops, which increase in size as the gas reaches higher densities. Gas which reaches the high density regime is highly molecular and takes a longer time to photodissociate back to lower H$_2$ fractions. Fig.~17 indicates that gas which reaches an H$_2$ fraction of 0.01 usually retains this fraction for only 10 to 20 Myr. However gas which goes on to obtain H$_2$ fractions of 0.5 retains this H$_2$ fraction for $\sim$40 Myr. This is clearly an upper limit for the expected cloud lifetime, as we do not include star formation and stellar feedback. The strong UV radiation and winds from newly formed massive stars would accelerate molecular cloud disruption.
\begin{figure}
\centerline{
\includegraphics[bb=50 350 650 780,scale=0.45]{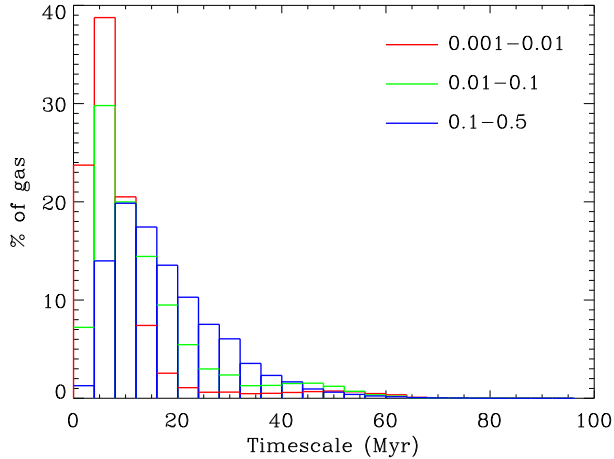}}
\caption{This histogram gives the distribution of timescales over which the gas reaches certain molecular gas fractions.
The timescales denote the time for the H$_2$ fraction of a particle to increase from 0.001 to 0.01, 0.01 to 0.1 and 0.1 to 0.5, as indicated.}
\end{figure} 

\begin{figure}
\centerline{
\includegraphics[bb=100 350 600 780,scale=0.45]{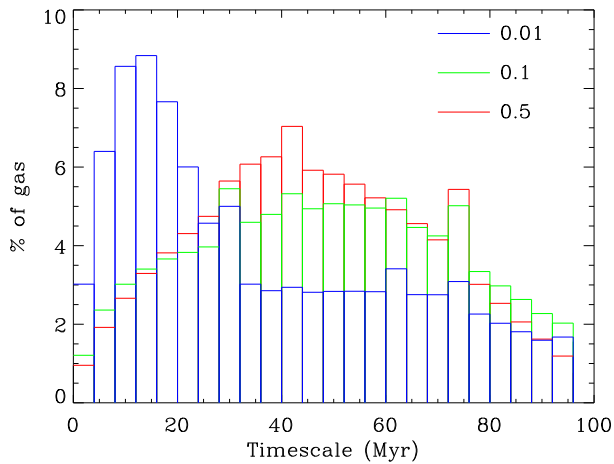}}
\caption{This figure shows the distribution of timescales gas remains over molecular gas fractions of 0.01, 0.1 and 0.5. Gas typically spends about 40 Myr with a molecular gas fraction of $>$ 0.5. This is however an upper limit on a molecular cloud lifetime, since stellar feedback is not included in these models.}
\end{figure}
\section{Discussion}

The calculations presented in this paper demonstrate that the atomic and molecular cooling discussed in
\citet{Glover2007a} permits the formation of molecular clouds in spiral shocks. Warm gas flowing into
the spiral potential is compressed, enabling it to cool to low temperatures, and initiating the rapid 
formation of H$_{2}$ in the cold gas. By `rapid' we mean that large quantities of H$_{2}$ form in
roughly a crossing time, consistent with the  ideas presented by a number of previous studies 
\citep{Vaz1996,Ball1999,Koyama2000,Elmegreen2000,Pringle2001}.

Our results differ from the two competing theories of GMC formation. In one, cloud formation is driven
by self-gravity (e.g. \citealt{Elmegreen1979}), and H$_{2}$ formation occurs on the cloud collapse timescale
(which may be significantly longer than the free-fall time for clouds that have a high level of
turbulent or magnetic support). In the other, GMC formation occurs through the coalescence of many small
cloudlets, which are dense but not self-gravitating \citep{Pringle2001}. Our
calculations cannot directly address the first of these  scenarios, since we do not include the effects
of the self-gravity of the gas. However, our results do highlight the importance of the gravitational
potential of the spiral arm for driving the formation of GMCs. Orbit crowding within the spiral arm
leads to cloud coalescence, as discussed in Section 3.2.5. The final picture is therefore similar to
that proposed by \citet{Pringle2001}.

The rapid formation of molecular clouds as suggested in the current study, as well as in a number of
more detailed chemical and dynamical calculations 
(e.g. \citealt{Ball1999,Ball2007,Glover2007a,Glover2007b, Hartmann2001, Heitsch2006,Heitsch2008,HAM2007, Henne2008,Vaz2006,Vaz2007}),
agrees well with current observational data. For example, it is consistent with the geometrically
derived fast star formation timescales in 14 nearby spiral galaxies observed with the 
VLT and the Spitzer infrared
satellite. \citet{Tamburro2008} measure the average angular offset between the H{\sc i} and 24$\,\mu$m
emissivity peaks along spiral arms and find typical timescales for the onset of star formation in this
sample of below 5 million years. Earlier studies on the position of dust emission, H{\sc i}, H{\sc ii},
and H$\alpha$ line from young stars measured across spiral shocks in M51 and M83 \citep{Allen1986,Tilanus1989,Leisawitz1989} come to similar
conclusions. Short cloud formation timescales are also consistent with the stellar age spread in
molecular clouds which exhibit a short peak of star-formation activity of 1 to 2 millions years duration
\citep{Hartmann2001,Hartmann2003}, possibly preceded by an acceleration period with very low activity of up to 10 million years  (e.g. \citealt{Palla2000,Palla2002}). Altogether, we conclude that a cloud formation
timescale of about 10 million  years or less as suggested by our calculations is consistent with
observational data \cite[see also ][]{Elmegreen2000, Elmegreen2007}.

Our study addresses another fundamental point of GMC formation: local self-gravity is not required to form
molecular clouds. The `clouds', or complexes of molecular gas, are simply a result of a change in
chemical state brought about by the processes described above. It is not necessary to assume that GMCs
are dominated by self-gravity. In particular, GMCs need not be in virial equilibirum, as previously
noted by \citet{Pringle2001}.  The path to molecular cloud formation is already underway at densities
of 0.1 to 1 cm$^{-3}$ and at temperatures of around 100 -- 300 K, the regime in which the self-shielding
becomes effective. The free-fall time at this density is around 50 Myr. Since this is longer than the
typical cloud formation timescale, it is unlikely that gravity can play a dominant role in the formation
of molecular clouds. Additionally, self-gravity has to overcome turbulence present in
the ISM, and tidal shear from the galaxy potential. Given these conditions, it seems likely that globally unbound GMCs can form, in which the star
formation efficiency would be naturally quite low \citep{Clark2005}.

\section{Conclusions}
The simulations presented in this paper indicate that cooling in spiral shocks leads to the formation of molecular clouds in the spiral arms of a galaxy. Self gravity of the gas is not required for this process. Cooling of the gas is very quick as gas enters the shock (almost instantaneous), hence cold gas clumps can agglomerate into larger structures, as described in previous isothermal simulations \citep{Dobbs2006,DBP2006,Dobbs2008}, within the spiral shock. Larger GMCs form after successive shocks, once both the arm and interarm material has become colder and denser. Again, similar to previous calculations, these clumps are sheared into interarm spurs.

For a disc of surface density comparable to the solar neighbourhood, around 70\% of the gas lies in the cold regime, and 20\% is molecular.
The transition from atomic to molecular gas is shown to be strongly dependent on the self shielding limit. Spiral shocks compress the gas sufficiently that large amounts of gas exceed the self shielding limit, thus enabling rapid formation of H$_2$. Gas obtains moderate H$_2$ fractions immediately after compression. The maximum H$_2$ fraction obtained then depends on the path of the gas through the shock. Gas which encounters little high density material may obtain a fraction of 0.01 or so for a short period (10 -- 20 Myr) after which photodissociation reduces the H$_2$ fraction. On the other hand gas may encounter dense, cold material, in which case it will undergo a stronger shock, be subject to greater cooling and become predominantly molecular. The time for the transition to fractions of 0.5 or so is typically 10 to 20 Myr, although this is a strict upper limit because our calculations cannot 
fully resolve internal turbulence and neglect self-gravity of the gas. True cloud formation timescales therefore will be considerably shorter. Without stellar feedback however, this gas tends to remain at high densities, and therefore retain appreciable H$_2$ fractions for substantially longer times ($\sim40$ Myr). We note the cloud formation time is generally not a well defined quantity, and suggest that determining the time from which gas enters a spiral arm, to the point at which it becomes observable (in CO), may be a more robust measure in future calculations.      

\section*{Acknowledgments}
We thank an anonymous referee for reviewing the paper. CLD also thanks Matthew Bate for assistance with numerical issues, and Jim Pringle for reading a draft of the paper.

This work, conducted as part of the award `The formation of stars and planets: Radiation hydrodynamical and magnetohydrodynamical simulations' made under the European Heads of Research Councils and European Science Foundation EURYI (European Young Investigator) Awards scheme, was supported by funds from the Participating Organisations of EURYI and the EC Sixth Framework Programme. This work was also carried out under the HPC-EUROPA project (RII3-CT-2003-506079), with the support of the European Community - Research Infrastructure Action under the FP6 "Structuring the European Research Area" Program). P.C.C. acknowledges support by the Deutsche
Forschungsgemeinschaft (DFG) under grant KL 1358/5 and via the
Sonderforschungsbereich (SFB) SFB 439, Galaxien im fr\"uhen Universum. Partial travel support was provided by the European Commission FP6 Marie Curie RTN CONSTELLATION (MRTN-CT-2006-035890). Calculations included in this paper were conducted on Exeter's Astrophysics Group SGI Altix ICE supercomputer, Zen, and the NEC Linux Cluster at the HLRS facility in Stuttgart.
Figures were produced using SPLASH, a visualisation package for SPH that is publicly available from http://www.astro.ex.ac.uk/people/dprice/splash/  \citep{splash2007}.
\appendix
\section{Resolution study}
To investigate the effects  of resolution, we repeated our fiducial calculation, with $\Sigma$= 10 \msp
 and $l_{\rm ph}=35$ pc using 1 and 4 million particles. The results for the 4 and 8 million particle calculations are very similar, although some discrepancies arise, particularly regarding the amount of cold and molecular gas, when there are only 1 million particles. 

On global scales, there is barely any difference in the structure of the disc for all the different resolution simulations. To highlight possible differences, Fig.~A1 zooms in on a section of the disc. The structure is very similar for the 4 and 8 million particle calculations. For the 1 million particle simulation, there is slightly less structure both along the spiral arms, and in the interarm spurs, since the small scale structure is not as well resolved. 

We also compared the thermal distribution of the gas, and the amount of H$_2$ formed at different resolutions. Figure~A2 indicates that the thermal distribution is very similar at different resolutions. The distributions mainly differ at low temperatures where the cold, dense regions are not as well resolved at lower resolutions. We compared the amount of molecular gas after 240 Myr in the 1, 4 and 8 million particle simulations, finding that 12, 21 and 22 \% of the gas in the disc is molecular. Thus with only 1 million particles the amount of molecular gas is significantly underestimated. Again the amount of molecular gas is similar in the 4 and 8 million simulations, although the 22\% for 8 million particles still represents a lower limit. 
\begin{figure}
\centerline{
\includegraphics[bb=40 0 560 700,scale=0.32,angle=270]{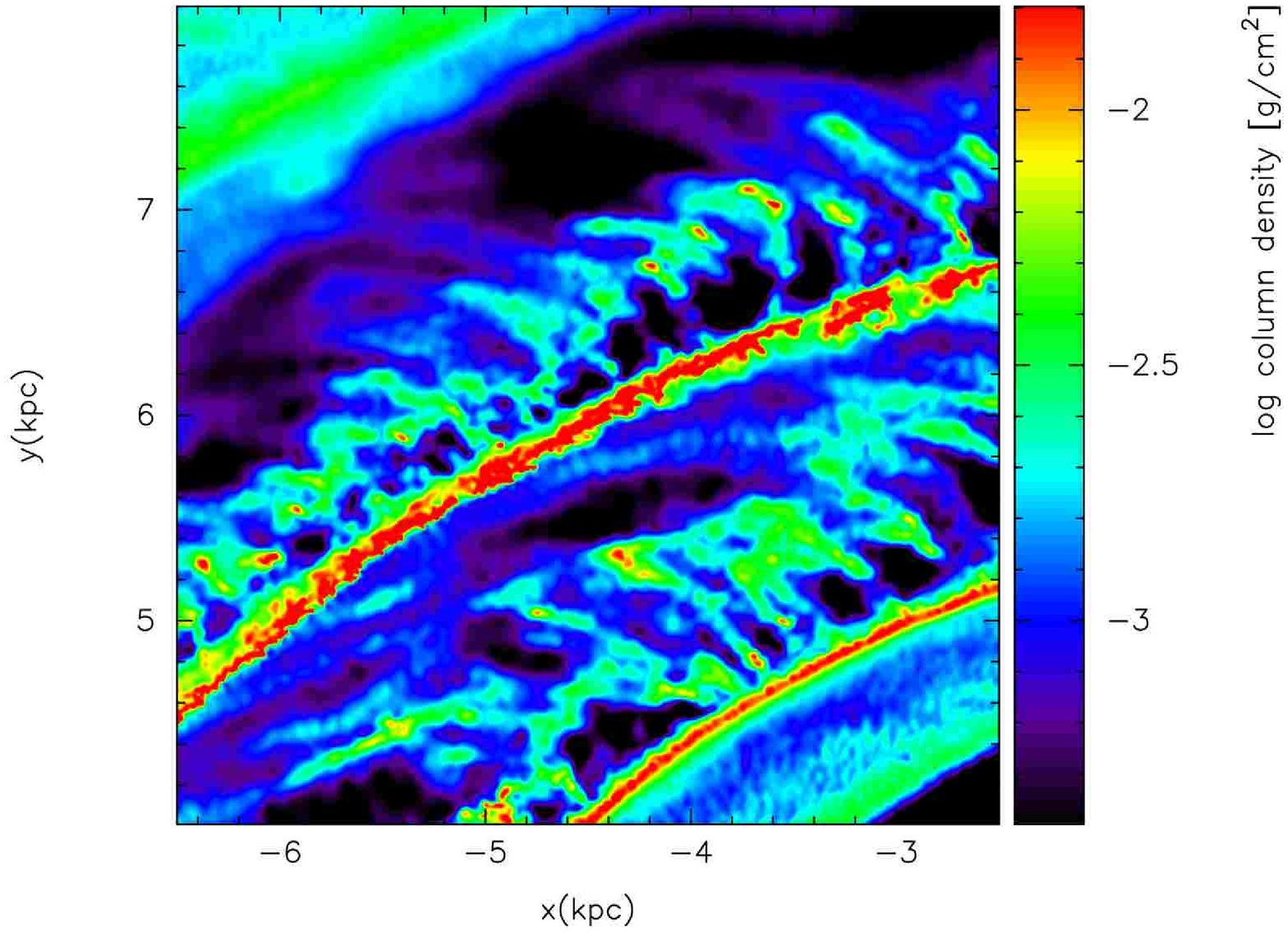}}
\centerline{
\includegraphics[bb=40 0 560 700,scale=0.32,angle=270]{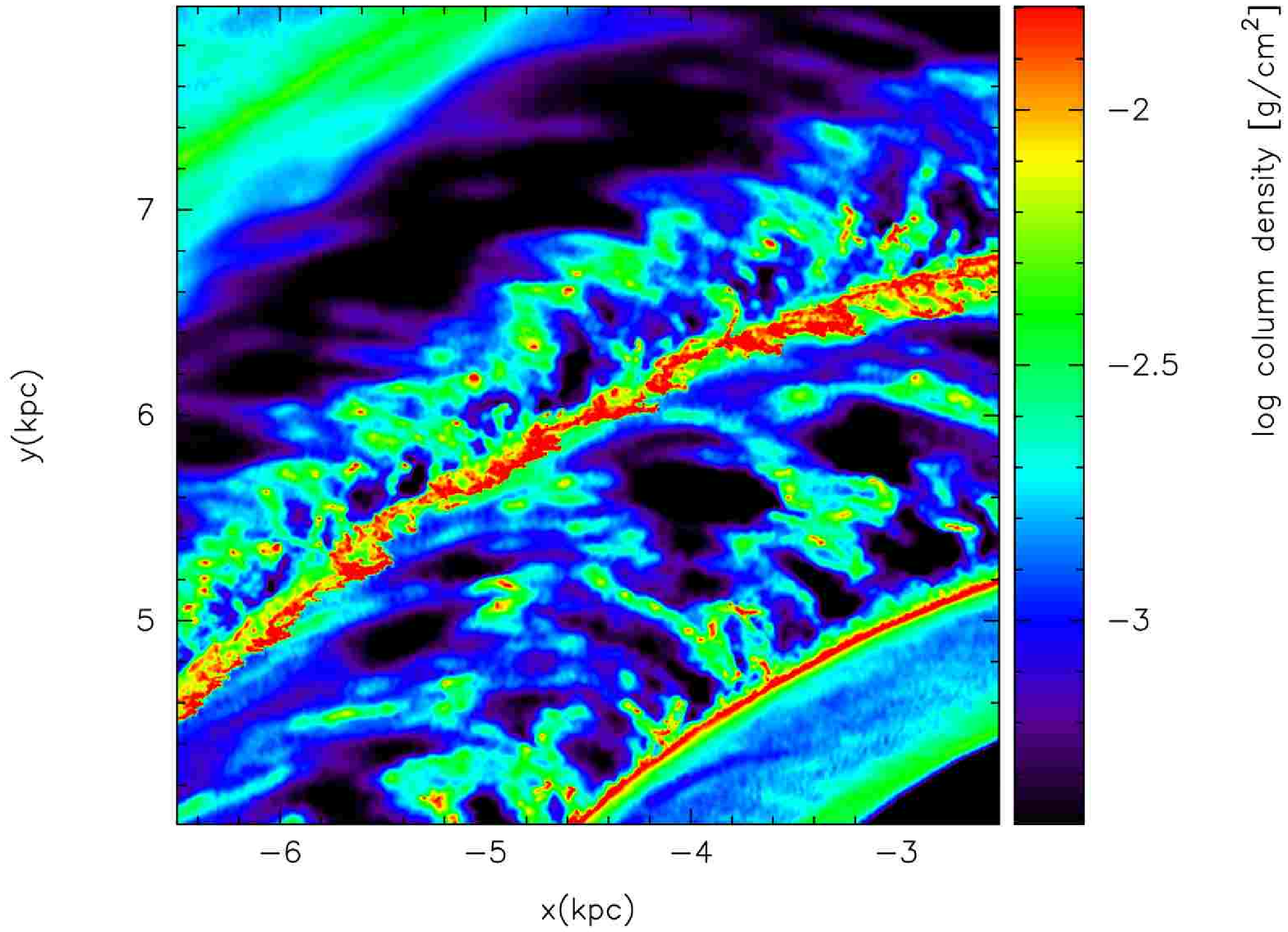}}
\centerline{
\includegraphics[bb=40 0 560 700,scale=0.32,angle=270]{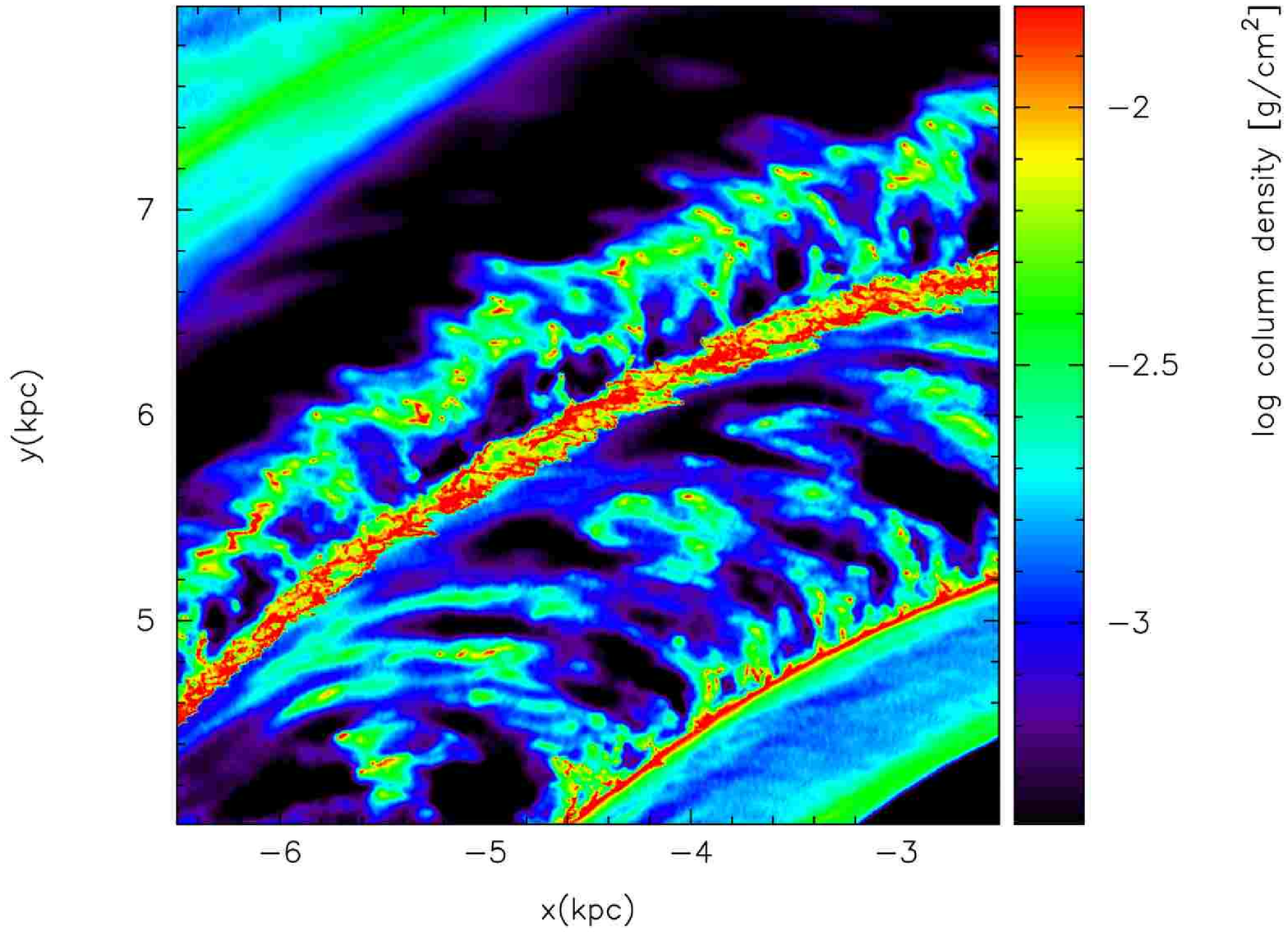}}
\caption{The column density of a section of the disc is displayed for the fiducial simulation, where $\Sigma$= 10 \mspnospace, with a resolution of 1 (top), 4 (middle) and 8 (lower) million particles. There is slightly less structure in the 1 million particle calculation, especially along the spiral arms. Generally however, the structure is very similar.}
\end{figure}
\begin{figure}
\centerline{
\includegraphics[bb=40 340 660 800,scale=0.43]{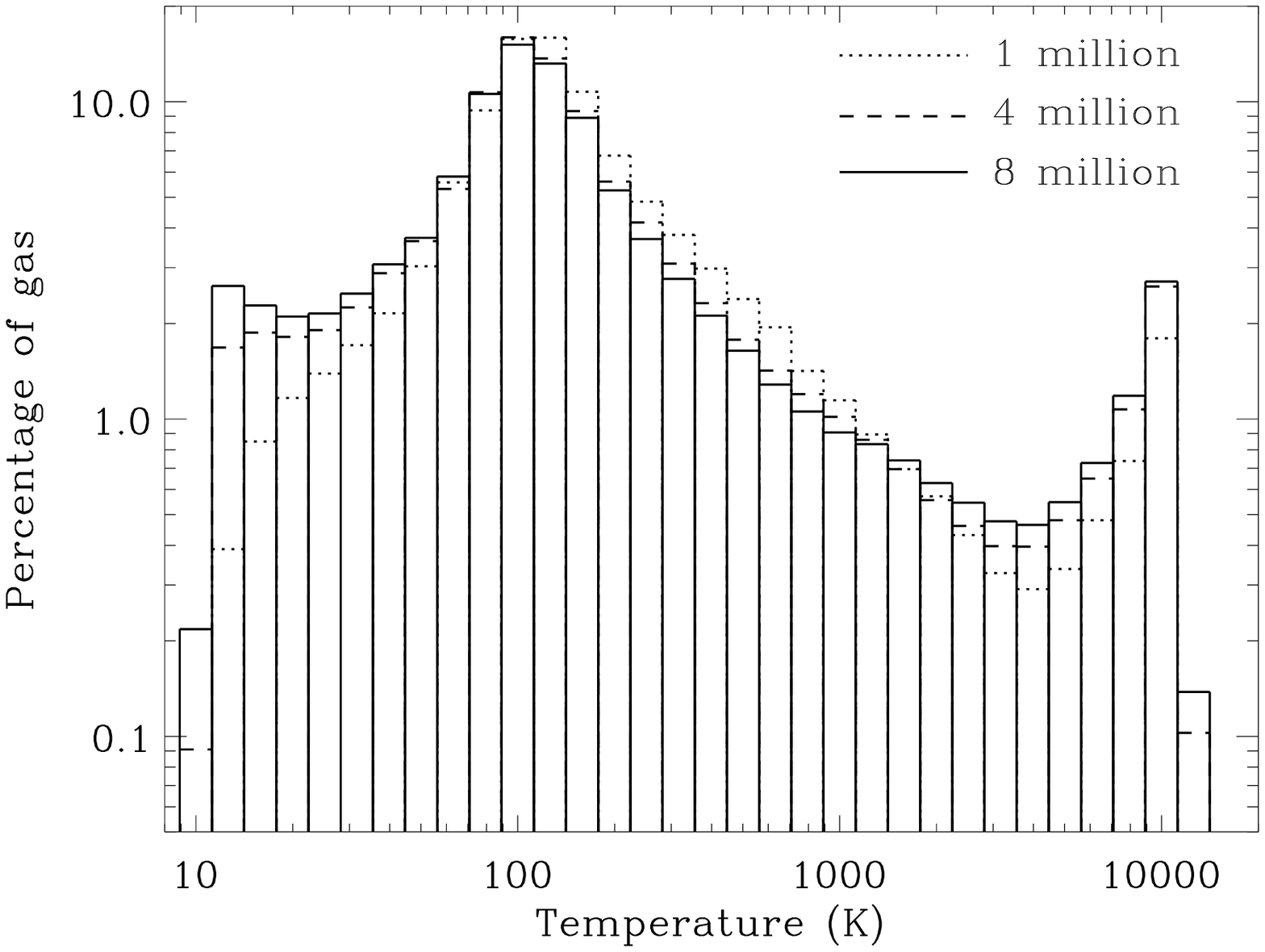}}
\caption{The temperature distribution is shown for the different resolution simulations. The distributions are very similar; they mainly deviate at low temperatures, where the high density regions are not as well resolved with only 1 million particles.}
\end{figure}
\bibliographystyle{mn2e}
\bibliography{Dobbs}

\bsp

\label{lastpage}

\end{document}